\documentclass[aps,pra, superscriptaddress]{revtex4-2}
\usepackage{amsmath,amssymb, graphicx,bm,braket}
\usepackage{caption}
\usepackage{mathrsfs}    

\usepackage[colorlinks=true, linkcolor=blue, citecolor=blue, urlcolor=blue]{hyperref}

\usepackage{subfig}
\begin{document}

\title{Toffoli and C$^\text{n}$NOT (n$>2$) gates  in a neutral-atom platform using Rydberg coupling and dark state resonances}

\author{Sinchan Snigdha Rej and Bimalendu Deb}
\affiliation{School of Physical Sciences, Indian Association for the Cultivation of Science, Jadavpur, Kolkata 700032, India. }

\date{\today}

\begin{abstract}
 We propose a protocol for realizing a Toffoli gate using neutral-atom qubits in optical tweezers. Two ground-state hyperfine levels of the atoms are considered as qubit states.  Our method relies on the strong and long-range interactions between atoms due to Rydberg excitations and the occurrence of dark states in the target qubit, with both control and target qubits being individually addressed with laser pulses. Our gate protocol enables precise control over the quantum states of individual qubits, effectively suppressing undesirable transitions to ensure high-fidelity gate performance. The gate fidelity is estimated to be about $96\%$ for realistic system parameters. We further demonstrate a C$^\text{n}$NOT gate with $n >2$ by exploiting the Rydberg antiblockade mechanism, which allows multiple atoms within the blockade radius to be simultaneously excited to the Rydberg states. Thus, our approach may open a promising route to multi-qubit controlled operations for quantum computation.
 
\end{abstract}

\maketitle

\section{Introduction}
Large-scale quantum information processing requires  highly efficient engineered systems for high-fidelity multi-qubit quantum gate operation. Among these systems, neutral atoms and trapped ions stand out as two of the most promising ones. The development of universal quantum gates has been a significant milestone in this endeavor. Notably, the  {Cirac-Zoller} gate \cite{PhysRevLett.74.4091,Monroe1995Demonstration} and the  {Mølmer–Sørensen} gate \cite{PhysRevA.62.022311,Sackett2000Experimental} were among the earliest demonstrations of universal quantum gates using trapped ions.

Subsequent to these advances in ion-based quantum computing, neutral atoms in optical tweezers have emerged as a highly promising platform for quantum computation over the past two decades, primarily due to their inherent scalability. A major step forward in neutral atom-based quantum computing was proposed by Jaksch et al. \cite{PhysRevLett.85.2208} and demonstrated experimentally later \cite{Urban2009Observation}. A key advantage in this approach lies in the ability to excite atoms to highly energetic states  known as Rydberg states characterized by a large principal quantum number. In these states, atoms experience strong and long-range interactions, governed either by dipole-dipole or van der Waals forces. A key effect that results  from these interactions is the Rydberg blockade. When a single atom within an ensemble is excited to a Rydberg state, the interaction-induced large shift of the Rydberg state of a nearby atom  prevents both the atoms from being excited to the same Rydberg state within a certain spatial extent, known as the blockade radius. This blockade mechanism plays a crucial role in the realization of high-fidelity two-qubit quantum gates \cite{PhysRevLett.104.010503, PhysRevA.92.022336, PhysRevA.94.032306, Shi2017RydbergGatesFree,PhysRevA.104.012615}. Depending on the Rydberg blockade mechanism,  numerous two-qubit gate protocols  have been developed, such as the use of global pulse \cite{levine2019parallel}, shaped and analytic pulses \cite{maller2017high}, dark states  \cite{PhysRevLett.102.170502,Petrosyan2017} and many more \cite{Liu2022,Shen2019,Guo2020,Xue2024,Zhao2018,Jin2024,Su2023RabiBlockade}. Recently, a hybrid platform has been proposed to implement the Rydberg blockade gate using ion-mediated interaction \cite{PhysRevA.110.062618}. In addition to the Rydberg blockade, another key mechanism, known as Rydberg antiblockade ({RAB}), offers an alternative approach to constructing quantum gates \cite{PhysRevA.95.022319, Wu2021AntiblockadeSWAP,Wu2021ResilientRydberg,  Li2023HighToleranceAntiblockade}. Unlike the blockade effect, the RAB allows multiple atoms within the blockade radius to be simultaneously excited to the Rydberg state under carefully controlled interactions and detunings. This phenomenon has been  explored as a  tool for multi-qubit gate implementation \cite{PhysRevA.95.022319}. 

Two-qubit gates must be decomposed into multi-layer circuits to implement conditional three-qubit operations, leading to increased circuit depth and error accumulation. Three-qubit Toffoli gate executes these operations directly, enabling more efficient syndrome extraction in stabilizer codes with a reduced number of ancillas and gate layers. This reduction in gate count and circuit depth is crucial for fault-tolerant quantum computation\cite{nielsen2000quantum} to improve the logical error rate\cite{Old2025fault_tolerant_three_qubit}.  Leveraging mechanisms like Rydberg blockade or antiblockade, native Toffoli gates can be realized with a minimal number of pulses, enabling fast, high-fidelity operations suitable for fault-tolerant quantum computation \cite{PhysRevA.95.022319, li2018onestep, su2019onestep,Khazali2020FastMultiqubit,Ashkarin2022ToffoliFineStructure,PhysRevApplied.18.034072}. Recently, a scheme for realizing a universal set of quantum gates based on  Rydberg blockade and  electromagnetically induced transparency (EIT) in atomic ensembles has been theoretically proposed \cite{PhysRevLett.102.170502} and experimentally demonstrated \cite{PhysRevLett.129.200501}. 

Here we propose two protocols for three-qubit Toffoli gate operation based on EIT and Rydberg blockade in a one or two-dimensional array of atomic qubits trapped in optical tweezers. In the first protocol, we consider a linear array  of qubits with sufficiently large inter-qubit spacing so that the qubits can be individually addressed with lasers. For the Toffoli gate, the target qubit is located at the midpoint between the two control qubits, as schematically shown in Fig.~\ref{F2}a. Two ground-state hyperfine levels of the atoms are considered as the qubit states.  The protocol is as follows: (i) Apply $\pi$ pulse on both control atoms simultaneously to drive the resonant transition from its qubit state $\ket{1}$ to the Rydberg state  $\ket{r}$. (ii)  Then, apply a pair of smooth Raman pulses, each with slowly time-varying amplitude on the target atom to drive a coherent two-photon transition between the two-qubit states   via an excited state $\ket{e}$ which is again resonantly coupled to a Rydberg state by another laser, leading to a pair of dark states for the target atom. (iii)  Finally, apply a second $\pi$ pulse simultaneously to both control qubits to complete the gate operation. In the second proposed protocol, we consider the three qubits arranged in a planar or two-dimensional configuration, as shown in Fig.~\ref{F3}a. The atoms are placed on the vertices of an equilateral triangular configuration where the spacing among the three atoms is the same. Here, we employ the two-atom RAB mechanism to excite both the control atoms to their respective Rydberg states simultaneously. The pulse sequence is the same as in the linear case, except that the control atoms are excited successively. Finally, we  generalize this protocol to construct a C$^\text{n}$NOT gate for $n>2$ in a two-dimensional atomic configuration of the tweezers using the RAB mechanism.  

We numerically calculate the population dynamics of the joint atomic states using qutip \cite{qutip2012} in order to validate our protocols. Our numerical results show that for high-fidelity three- or multi-qubit gate operations, it is crucial to choose the parameters of the pulses appropriately so that the system remains close to the dark states during the gate operation for adiabatic transfer of the populations.  We have calculated the gate fidelity using the trace-preserving, quantum-operator-based formula \cite{nielsen2002simple}. The fidelity is found to be about $96\%$ for the Toffoli gate and $94\%$ for the C$^\text{3}$NOT gate for the realistic system parameters. We have considered spontaneous decay of the Rydberg states as the source of gate errors.

The paper is organized in the following way: In section \ref{LT} we present and demonstrate our protocol for the Toffoli gate in a linear configuration of atomic qubits. In section \ref{PT}, we generalize this  protocol of the Toffoli gate for a planar triangular configuration using RAB and EIT.  We then construct a C$^3$NOT gate using the same protocol in section \ref{NT}. We present our numerical results and detailed discussions in section \ref{RD}.

\section{\label{S2A} Toffoli gate}
\subsection{\label{LT}Linear atomic configuration}
\begin{figure} [t]
     \includegraphics[width=15 cm]{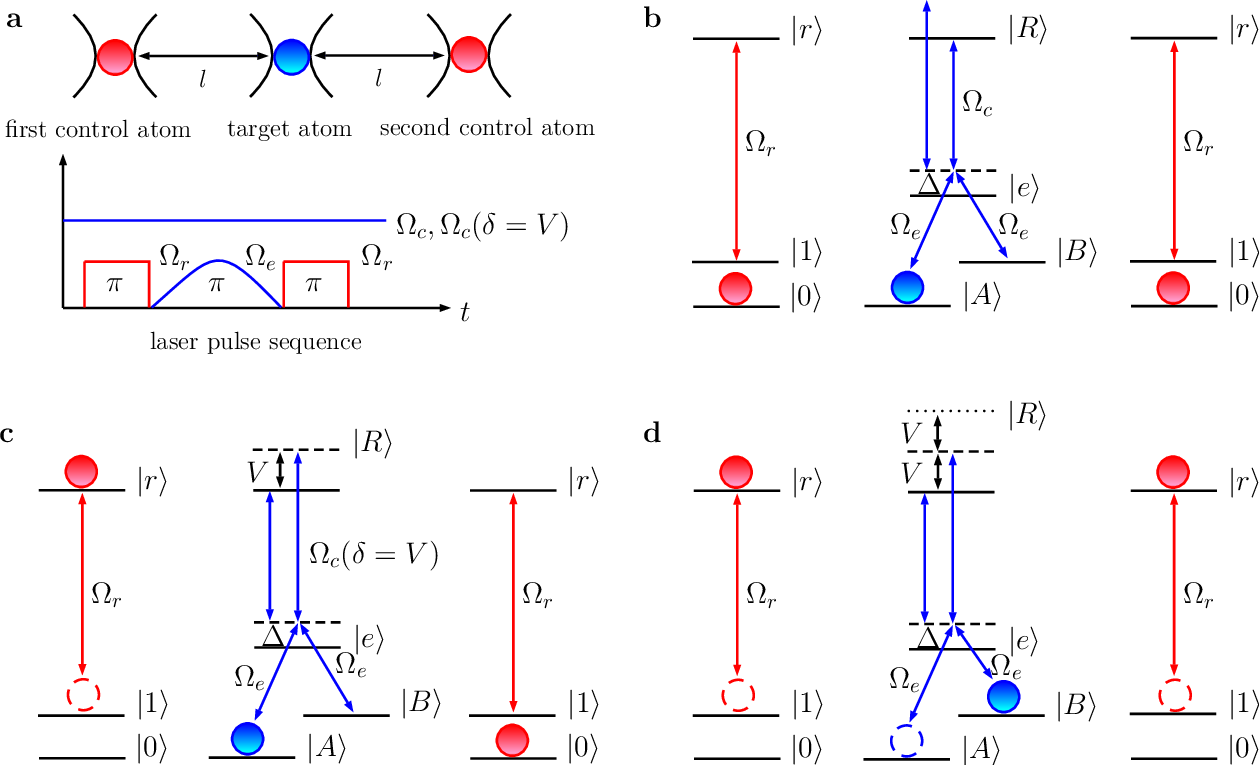}\\
    \caption{\label{F2} (a) Two control atoms and one target atom, configured in a linear arrangement with an inter-atomic separation $l$ and the atoms are trapped in three separate optical tweezers.  The laser pulse sequence for the Toffoli gate is shown, where the term in the first bracket after $\Omega_c$ indicates the value of detuning $\delta$ of the laser, with $V$ being the control-target atom interaction. (b) When both control qubits are initially at $\ket{0}$, the respective Rydberg states $\ket r$ remain unpopulated after the first $\pi$ pulse resulting in no shift in the target atom's Rydberg state $\ket{R}$ that is coupled with $\ket{e}$ with the laser of Rabi frequency $\Omega_c$, and after the Raman $\pi$ pulse is applied, the target qubit remains unchanged since it is prepared in a dark state (see text). (c) When either of the control atoms is initially at $\ket{1}$ and the other one is at $\ket{0}$ the first $\pi$ pulse will excite the atom in $\ket 1$ to $\ket{r}$ resulting in the shift of $\ket{R}$ by $V$. The shifted Rydberg state of the target atom is coupled with $\ket{e}$ with another laser of the same Rabi frequency $\Omega_c$ but with a detuning $\delta=V$. As a result, there is again no change in the target qubit due to the dark state condition. (d) When both control atoms are at $\ket{1}$, simultaneous $\pi$ pulses will excite both atoms to $\ket{r}$ leading to the shift of $\ket{R}$ by $2V$ that is large enough to break the EIT condition enabling the transformation $\ket{A}\longleftrightarrow\ket{B}$ adiabatically due the Raman $\pi$ pulse. } 
\end{figure}
To implement the Toffoli gate, we arrange two control atoms and one target atom in a linear configuration, as illustrated in Fig.~\ref{F2}a. The ground-state sub-levels (qubit states) of the control atoms are denoted as $\ket{0}$ and $\ket1$, and for the target atom $\ket A$, $\ket{B}$ and $\ket{e}$ denote the two qubit states and the excited intermediate state, respectively. The Rydberg states of the control atoms and target one are denoted as $\ket r$ and $\ket R$, respectively. The atoms are optically trapped, maintaining a fixed control-target atom distance, denoted by $l$. The qubit state $\ket{1}$ of the control atom is resonantly coupled to  $\ket{r}$ by a laser with Rabi frequency $\Omega_r$. For the target atom, a pair of smooth Raman pulses, each with slowly time-varying Rabi frequency $\Omega_e(t)$ which has an amplitude $\Omega_e$, drives a coherent two-photon transition between $\ket{A}$ and $\ket{B}$ via $\ket{e}$, which has lifetime $\gamma_e^{-1}$. The states $\ket A$ and $\ket B$ are off-resonantly coupled with $\ket e$ with a blue detuning $\Delta \gg \gamma_e$ to minimize spontaneous emission losses from $\ket e$. A separate laser ($\mathcal{L}_1$), characterized by a Rabi frequency $\Omega_c$, couples $\ket{e}$ to $\ket{R}$. The system is designed such that the  states $\ket{A}$ and $\ket{B}$ are in two-photon resonance with $\ket{R}$, ensuring efficient population transfer. However, in the pulse sequence, an additional laser ($\mathcal{L}_2$) with Rabi frequency $\Omega_c$ and a detuning  $\delta$ is included. The parameters satisfy the condition $\Delta \gg \Omega_c > \Omega_e$, which allows controlled excitation dynamics. The pulse sequence for implementing the Toffoli gate is illustrated in Fig.~\ref{F2}a. The sequence begins with simultaneous $\pi$ pulses applied to the control qubits. This is followed by a pair of smooth adiabatic Raman $\ pi$-pulses on the target qubit, characterized by a $\Omega_e(t)$ that satisfies $\int_0^T dt \Omega_e^2(t)/2\Delta  = \pi.$ Finally, another simultaneous $\pi$ pulses are applied to the control qubits, completing the gate operation.

 In the first scenario, when both control qubits are initially in the state $\ket{0}$, the Rydberg states of both the control atoms remain unpopulated. Consequently, the Hamiltonian of the target atom is as follows
 \begin{equation}  
     \mathcal{H}^1(t,\delta)=  \Omega_e(t)/2(\ket{A}\bra{e}+\ket{B}\bra{e})+ \Omega_c/2(\ket{e}\bra{R}+ \ket{e}\bra{R}e^{i\delta t}) - \Delta\ket{e}\bra{e} + \text{H.c.}.
     \label{Ht1}
 \end{equation}

 The effective Hamiltonian after the Magnus expansion \cite{brinkmann2016introduction} up to second order is given by
 \begin{equation}  
 \begin{aligned}
     \mathcal{H}_{eff}^1(t,\delta)=&  \Omega_e(t)/2(\ket{A}\bra{e}+\ket{B}\bra{e})+ \Omega_c/2(\ket{e}\bra{R}+ \ket{e}\bra{R}(e^{i\delta t/2}\sin({\delta t/2)/(\delta t/2)}) )\\&- \Delta\ket{e}\bra{e} +\Omega_c^2/4(1/\delta-\sin{\delta t}/(\delta^2t))(\ket{e}\bra{e}-\ket{R}\bra{R})+ \text{H.c.},
     \end{aligned}
 \end{equation}
 for $\delta\gg\Omega_c$ the time dependent terms will contribute negligibly and ignoring these terms we calculate the two dark states $\ket{\mathcal D_1}=1/\sqrt{2}(\ket{A}-\ket{B)}$ and $\ket{\mathcal D_2}=(1+y^2)^{-1/2}[1/\sqrt{2}(\ket{A}+\ket{B}-y\ket{R})]$ where $y=\sqrt{2}\Omega_e(t)/\Omega_c $. Satisfying the condition $\Omega_c/\Omega_e>2 $, the target qubit follows the dark state $\ket {\mathcal D} =\frac{1}{\sqrt2}(\ket {\mathcal{D}_1} +\ket {\mathcal{D}_2)} $ during evolution, resulting in the transformation $\ket{00A} \longrightarrow \ket{00A}$,
as depicted in Fig.~\ref{F2}b. 

Now, if one of the control qubits is initially in the state $\ket{1}$, it transitions to $\ket{r}$ due the first $\pi$ pulse, while the other control qubit remains at $\ket{0}$. Since one control atom is in the state $\ket{r}$, the Rydberg state $\ket{R}$ of the target atom experiences a shift $V$ due to the Rydberg-Rydberg interaction. This shift would normally break the EIT condition, but we take an additional laser $\mathcal{L}_2$ with the same Rabi frequency $\Omega_c$ but with a blue detuning $\delta$ to couple the shifted Rydberg state to the intermediate state $\ket{e}$ resonantly. The Hamiltonian for the target atom will be as follows,
\begin{equation}
     \mathcal{H}^2(t,\delta,V)= \mathcal{H}^1(t,\delta) + V\ket{R}\bra{R} 
     \label{eq2}
 \end{equation}
 After rotating with respect to $e^{-iV\ket{R}\bra{R}}$ we get the Hamiltonian as follows
 \begin{equation}
     {\mathcal{H}^2}'(t,\delta,V)=  \Omega_e(t)/2(\ket{A}\bra{e}+\ket{B}\bra{e})+ \Omega_c/2(\ket{e}\bra{R}e^{-iVt}+\ket{e}\bra{R}e^{i(\delta-V)t}) - \Delta\ket{e}\bra{e} + \text{H.c.}.
     \label{eq2}
 \end{equation}
 After the Magnus expansion and fulfilling the condition $\delta=V\gg\Omega_c$ we get two dark states as in the previous case. During the smooth Raman pulse, the target qubit again follows the dark states, provided that the condition $\Omega_c / \Omega_e>2$ is satisfied. Consequently, the transformation $\ket{10A} \longrightarrow \ket{10A}$ is fulfilled,
as depicted in Fig.~\ref{F2}c.

In the third scenario, when both control qubits are initially in the state $\ket{1}$, they transition to their respective Rydberg states $\ket{r}$ following the first simultaneous $\pi$ pulses. The distance between the two control atoms is adjusted such that the interaction between them is negligible. However, the target atom experiences interactions with both control atoms. Consequently, the state $\ket{R}$ of the target atom experiences a shift of $2V$. 
Since no laser exists to couple the shifted Rydberg state to this intermediate state $\ket{e}$, the Hamiltonian for the target atom modifies to 
\begin{equation}
    {\mathcal{H}^3}(t,V) = {\mathcal{H}^2}'(t,\delta=V) + V\ket{R}\bra{R}.
\end{equation}
This shift effectively breaks the EIT condition when $V >    \Omega_c^2 / (4\Delta)$, thereby enabling the target qubit to transition between $\ket{A} \longleftrightarrow \ket{B}$ during the Raman $\pi$ pulse. Upon applying the final $\pi$ pulse on both control qubits simultaneously, we achieve the desired transformation
$\ket{11A} \longleftrightarrow \ket{11B}$ (Fig.~\ref{F2}d).

\subsection{\label{PT} Planner atomic configuration}
\begin{figure} 
     \includegraphics[width=4.0cm]{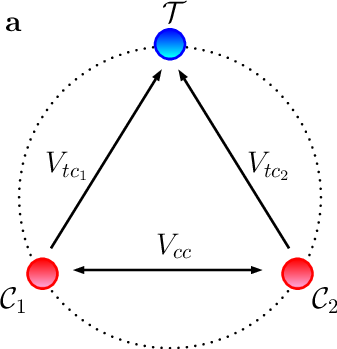} \hspace{10mm}
     \includegraphics[width=8.5cm]{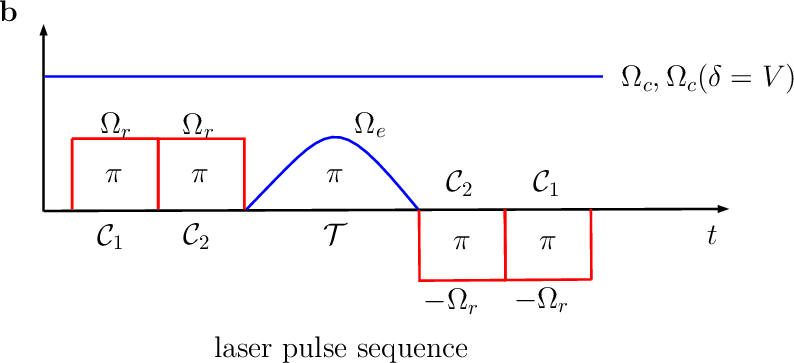}\\
     \includegraphics[width=5cm]{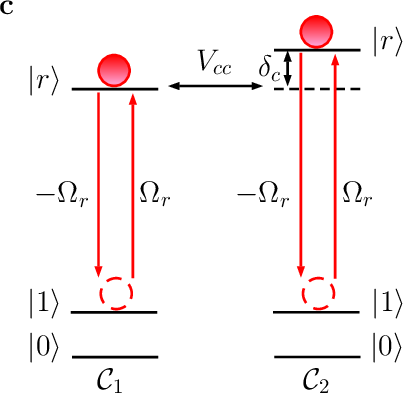}
     \includegraphics[width=9cm]{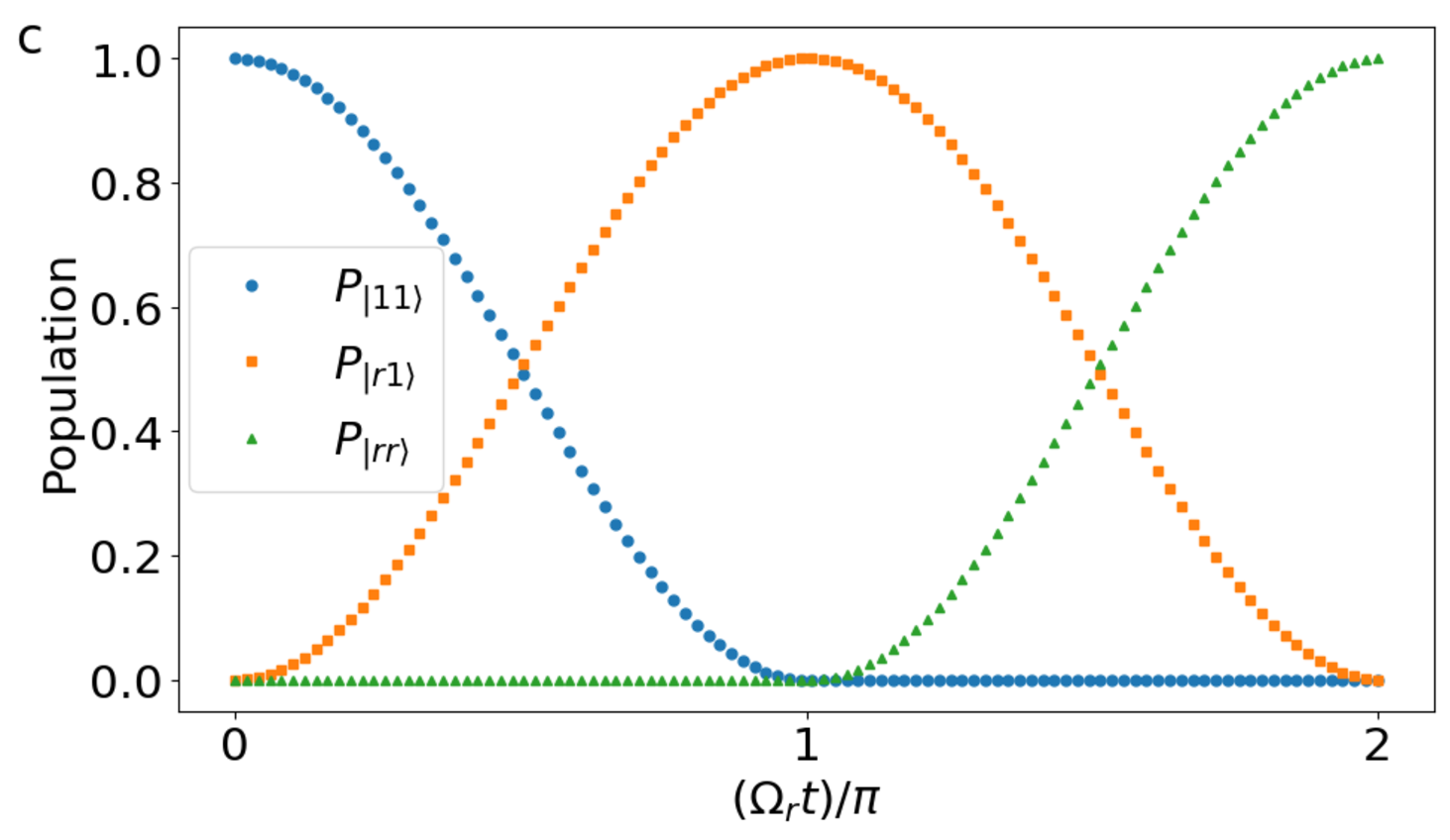}
     \\
    \caption{\label{F3}(a) Atomic arrangement to implement the Toffoli gate in 2D configuration where $\mathcal{C}_i$ and $\mathcal{T}$ denote the control and target atoms, respectively, $V_{cc}$ is the interaction between two control atoms and $V_{tc_i}$ is the interaction between the target and the $\mathcal{C}_i$ control atom.
    (b) Laser pulse sequence for the Toffoli gate where the term in the first bracket after $\Omega_c$ indicates the value of detuning of the laser.
    (c) Pulses are applied on the control qubits with appropriate detuning to fulfil the RAB condition and complete the gate protocol.
    (d) Population transfer from $\ket{11}$ to $\ket{rr}$ state using the two-atom RAB mechanism. } 
\end{figure}
We can face a major hurdle while scaling up this system in a linear atomic arrangement. The control-target atom distances for each control atom may differ, and as a result, all control atoms will not interact with the target atom equally. To overcome this issue, we arrange all the atoms in a two-dimensional plane, keeping both control atoms at equal distances from the target atom. In this scenario, the control atoms might get closer to each other than the linear atomic configuration, and they might interact with each other. To counteract control-control atom interaction, we can use the RAB mechanism. 

   Here we discuss the procedure to construct a Toffoli gate in a two-dimensional atomic arrangement. We arrange the target atom and two control atoms in an equilateral triangular configuration, as shown in Fig.~\ref{F3}a, where the distance between the control atoms is the same as the control-target atom distance, hence $V_{tc_i}=V_{cc}$, where $V_{tc_i}$ and $V_{cc}$ indicate the control-target atom interaction and control-control atom interaction. We utilize the two-atom RAB mechanism to simultaneously populate the Rydberg states of both control atoms within the blockade radius. The gate protocol remains the same as before for the initial states $\ket{00}$ and $\ket{10}$.
   To achieve the two-atom RAB condition, we have to apply a largely detuned laser on the second control atom. As a result, $\ket{0r}$ will never be populated from the initial state $\ket{01}$, unlike the linear case. In this scenario, following the Hamiltonian $\mathcal{H}^1(t,\delta)$, the EIT condition will be maintained. However, when both control atoms are initially in $\ket{11}$, $\ket{rr}$ can be populated, as we explain below.

 Let us now examine the two-atom RAB mechanism with sequential driving in detail. First, we apply a resonant $\pi$ pulse to the first control atom, inducing the transition $\ket{11} \longrightarrow -i\ket{r1}$ under the influence of the Hamiltonian
\begin{equation}
    {H}^2_1 =    \Omega_r/2(\ket{r1}\bra{11} + \text{H.c.}).
\end{equation}
Secondly, a dispersive $\pi$ pulse is applied to the second control atom with a blue detuning $\delta_c$. The system evolves under the following Hamiltonian
\begin{equation}
     {H}^2_2 =  \Omega_r/2 \left( \ket{rr}\bra{r1} e^{-i \delta_c t} + \text{H.c.} \right) + V_{cc} \ket{rr}\bra{rr},
\end{equation}
where $V_{cc}$ represents the interaction potential between the two control atoms. By transforming to a rotating frame with respect to $e^{-i V_{cc} \ket{rr}\bra{rr} t}$, the Hamiltonian becomes
\begin{equation}
     {{H}^2_2}' =  \Omega_r/2\left( \ket{rr}\bra{r1} e^{-i (\delta_c - V_{cc}) t} + \text{H.c.} \right).
\end{equation}

Under the condition $\delta_c = V_{cc}$, the system undergoes the transition $\ket{r1} \longrightarrow -i\ket{rr}$, populating the Rydberg states of both control atoms, as illustrated in Fig.~\ref{F3}d. Once the $\ket{rr}$ state is populated, the target atom experiences a shift of $2V$ ( $V_{tc_1} = V_{tc_2} = V$) in the Rydberg state $\ket{R}$ of the target atom. Due to the absence of a coupling laser between the shifted Rydberg state and the intermediate state $\ket{e}$, the EIT condition is broken, allowing transformation $\ket{A} \longleftrightarrow \ket{B}$ to occur. 

To complete the protocol, we apply the reverse operations of the second and first steps, respectively, resulting in the desired transformations.
The pulse sequence and the steps described above are illustrated in Fig.~\ref{F3}b and Fig.~\ref{F3}c, respectively.

\section{\label{NT} C$^3$NOT gate  }
 \begin{figure}[!h] 
\centering
     \includegraphics[width=4.0cm]{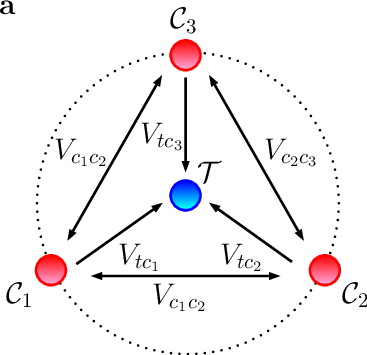}
     \includegraphics[width=10cm]{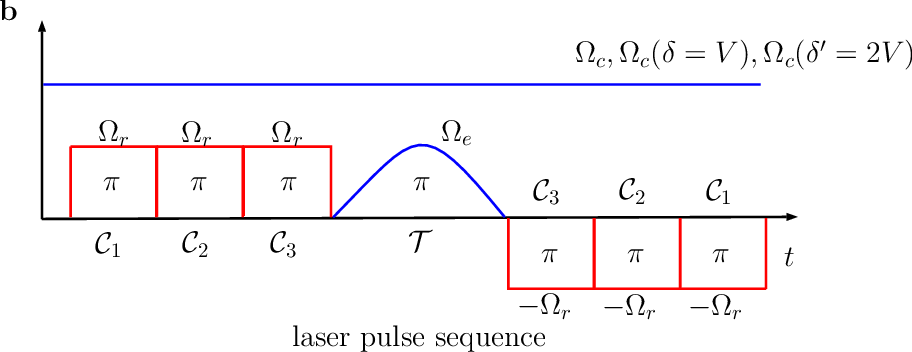}\\
     \vspace{0.5cm}
     \includegraphics[width=5.0cm]{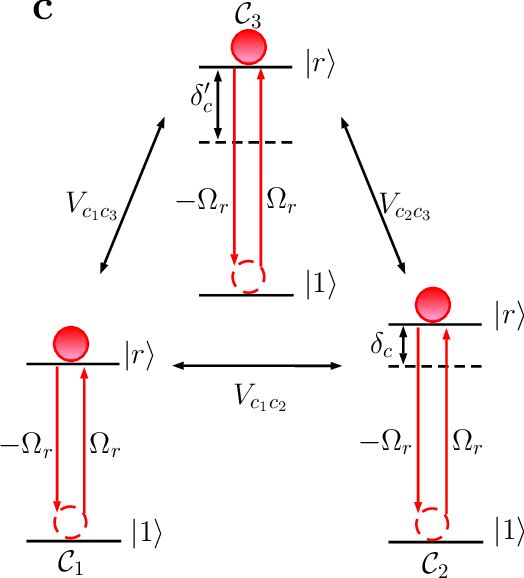}\hspace{1cm}
     \includegraphics[width=9cm]{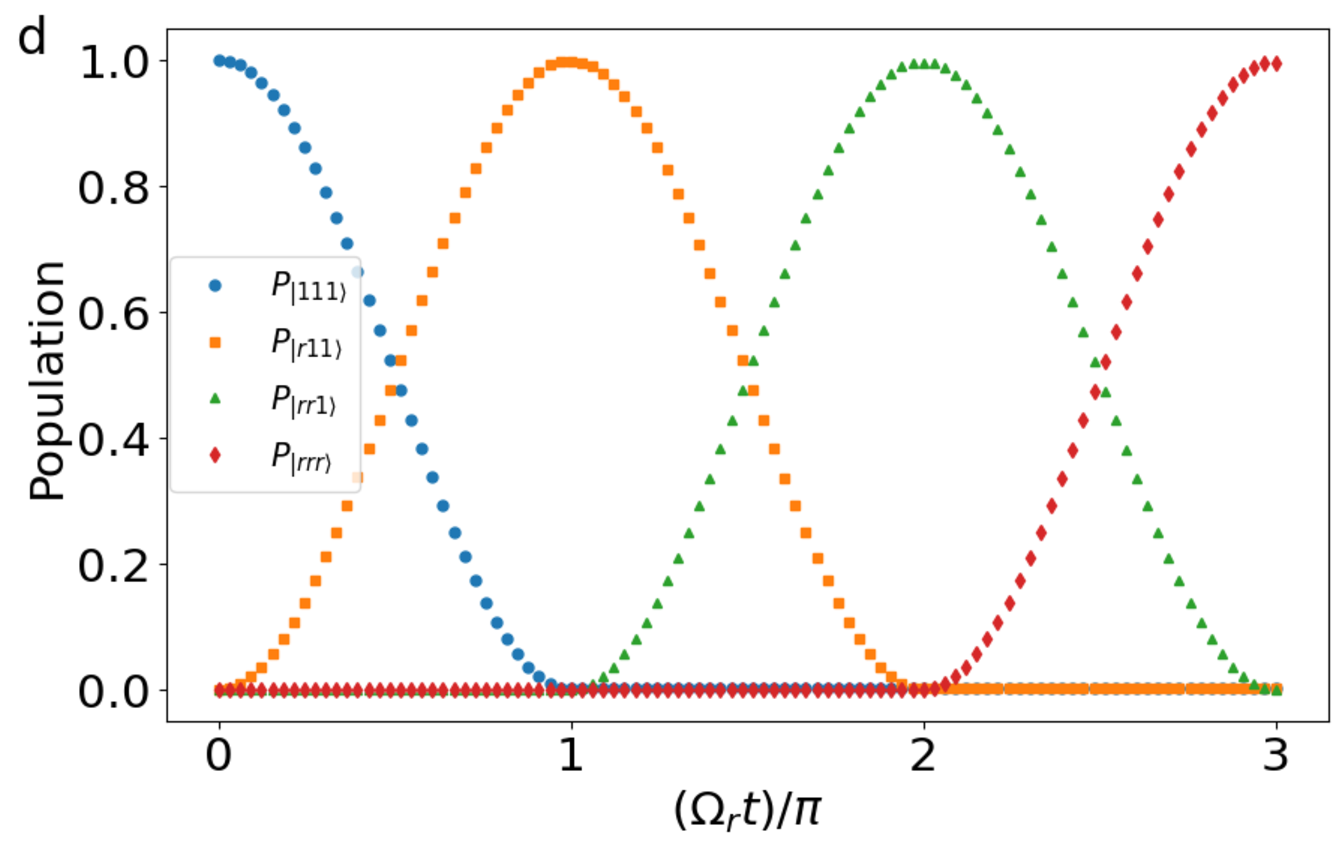}
     \\
    \caption{\label{F4}(a) Atomic arrangement to implement the C$^3$NOT gate in a 2D configuration.
    (b) Laser pulse sequence for the C$^3$NOT gate where the terms in the first bracket after $\Omega_c$ indicate the value of detuning of the laser.
    (c) Pulses are applied on the control qubits to get the RAB regime and complete the gate protocol.
    (d) Population transfer from $\ket{111}$ to $\ket{rrr}$ using the three-atom RAB formalism.}    
\end{figure}
To implement the C$^3$NOT gate, we arrange the three control atoms at the three vertices of an equilateral triangle, and the target qubit is at the centre of this triangle as shown in Fig.~\ref{F4}a to achieve RAB for all three control atoms. Let us arrange the initial states of control atoms in four different sets for better understanding; first set \{$\ket{000},\ket{010},\ket{001},\ket{011}$\}; second set \{$\ket{100},\ket{101}$\}; third set \{$\ket{110}$\}; fourth set \{$\ket{111}$\}. Here we will apply one resonant pulse on the first atom and two largely detuned laser pulses on the second and third atoms to couple $\ket1$ and $\ket r$. This will prevent population transfer from the states in the first set but the states in the second set will be transformed to \{$\ket{r00},\ket{r01}$\} respectively. The state in the third set will be transformed into $\ket{rr0}$ following the two-atom RAB mechanism. Here we introduce another laser $\mathcal{L}_3$ with Rabi frequency $\Omega_c$ and detuning $\delta'$ to couple the new shifted Rydberg state to $\ket e$ of the target atom. 

For the first set of initial states of the control atoms, the target atom will follow the Hamiltonian. 
\begin{equation} \mathscr{H}^1(t,\delta,\delta')=\mathcal{H}^1(t,\delta)+\Omega_c/2\ket{e}\bra{R}e^{i\delta't} + \text{H.c.}.
\end{equation}
Again, by doing the Magnus expansion of the above Hamiltonian and satisfying $\delta,\delta'\gg\Omega_c$, we can get an effective Hamiltonian where the time-dependent terms will contribute negligibly. Just like the previous case, here we get two dark states to maintain the EIT condition while satisfying the previously mentioned criteria.
The second set of initial states will follow the Hamiltonian 
\begin{equation}
\mathscr{H}^2(t,\delta,\delta',V)=\mathcal{H}^2(t,\delta,V)+\Omega_c/2\ket{e}\bra{R}e^{i\delta't}  + \text{H.c.} 
\end{equation}
and after rotating with respect to $e^{-iV\ket{R}\bra{R}}$ we get 
\begin{equation}
    {\mathscr{H}^2}'(t,\delta,\delta',V)={\mathcal{H}^2}'(t,\delta,V)+\Omega_c/2\ket{e}\bra{R}e^{i(\delta'-V)t}+ \text{H.c.}.
\end{equation}
Fulfilling the condition $\delta=V\gg\Omega_c$ and $(\delta'-V)\gg \Omega_c$, here we can also achieve the blockage of population transfer from the ground states of the target atom. For the third set, the Hamiltonian, the target atom will follow is 
\begin{equation}
    \mathscr{H}^3(t,\delta',V)={\mathscr{H}^2}'(t,\delta=V,\delta',V)+V\ket{R}\bra{R}
\end{equation}
and a rotating frame transformation with respect to $e^{-iV\ket{R}\bra{R}}$ will give 
\begin{equation}
\begin{aligned}
    {\mathscr{H}^3}'(t,\delta',V) =&\Omega_e(t)/2(\ket{A}\bra{e}+\ket{B}\bra{e})+ \Omega_c/2(\ket{e}\bra{R}e^{-i(2V)t}+\ket{e}\bra{R}e^{-iVt}+\ket{e}\bra{R}e^{i(\delta'-2V)t})
    \\&- \Delta\ket{e}\bra{e} + \text{H.c.}.
\end{aligned}
\end{equation}
When $\delta'=2V$ the shifted state $\ket{R}$ becomes resonantly coupled to $\ket{e}$, and for $V\gg\Omega_c$, neglecting the time-dependent terms from the effective Hamiltonian, we satisfy the EIT condition. Consequently, no transition occurs between $\ket{A}$ and $\ket{B}$.

Let us now consider the case where all three control atoms are initially in the state $\ket{1}$. For simplicity, we take all three control atoms at equal distance from each other so that $V_{c_1c_2}=V_{c_1c_3}=V_{c_2c_3}=V_{cc}$.
In the first step, a resonant $\pi$-pulse is applied to the first control atom, inducing the transition $\ket{111} \longrightarrow -i\ket{r11}$, as governed by the Hamiltonian:

\begin{equation}
     {H}^3_1 = \frac{\Omega_r}{2} (\ket{r11}\bra{111} + \text{H.c.}).
\end{equation}
In the second step, a dispersive $\pi$-pulse is applied to the second control atom with a blue detuning $\delta_c$, driving the system under the Hamiltonian:

\begin{equation}
    {H}^3_2 = \frac{\Omega_r}{2} \left( \ket{rr1}\bra{r11} e^{-i \delta_c t} + \text{H.c.} \right) + V_{cc} \ket{rr}_{12}\bra{rr} \otimes \mathcal{I}_3.
\end{equation}
Transforming to the rotating frame with respect to the unitary $e^{-iV_{cc} \ket{rr}_{12} \bra{rr} \otimes \mathcal{I}_3}$ and setting the detuning $\delta_c = V_{cc}$, the effective Hamiltonian simplifies to:

\begin{equation}
     {{H}^3_2}' = \frac{\Omega_r}{2} \left( \ket{rr1}\bra{r11} + \text{H.c.} \right),
\end{equation}
which results in the transformation $\ket{r11} \longrightarrow -i\ket{rr1}$.  
In the third step, another dispersive $\pi$-pulse with blue detuning $\delta_c'$ is applied to the third control atom. The corresponding Hamiltonian for the system is:

\begin{equation}
\begin{aligned}
     {{H}^3_3} = & \frac{\Omega_r}{2} \left( \ket{rrr}\bra{rr1} e^{-i \delta_c' t} + \text{H.c.} \right) + V_{cc} \ket{rr}_{12} \bra{rr} \otimes \mathbb{I}_3 \\
    & + V_{cc} \ket{rr}_{13} \bra{rr} \otimes \mathbb{I}_2 + V_{cc} \ket{rr}_{23} \bra{rr} \otimes \mathbb{I}_1.
\end{aligned}
\end{equation}
Since the term $V_{cc} \ket{rr}_{12}\bra{rr} \otimes \mathbb{I}_3$ commutes with the rest of the Hamiltonian, it can be treated independently. By moving into the rotating frame with respect to the remaining interaction terms and choosing $\delta_c'= 2V_{cc}$, we realize the three-atom RAB condition (see Fig.~\ref{F4}c), enabling the transformation $\ket{rr1} \longrightarrow -i\ket{rrr}$ under the effective Hamiltonian:

\begin{equation}
     {{H}^3_3}' = \frac{\Omega_r}{2} \left( \ket{rrr}\bra{rr1} + \text{H.c.} \right).
\end{equation}
At this stage, the target atom experiences a Rydberg energy shift of $3V$ $(V_{tc_1}=V_{tc_2}=V_{tc_3}=V)$ due to all three control atoms being in the Rydberg state. This shift will necessarily break the EIT condition following the Hamiltonian 
\begin{equation}
{\mathscr{H}^4}(t,V)={\mathscr{H}^3}'(t,\delta'=2V,V)+V\ket{R}\bra{R}.    
\end{equation}
Due to the absence of any coupling laser between the new shifted state $\ket R$ and $\ket e$, allowing for the required transformation $\ket A\longleftrightarrow\ket{B}$. 
To complete the gate protocol, we reverse the third, second, and first steps sequentially to implement the C$^3$NOT gate. The same procedure can be followed to construct an n-atom RAB condition for realizing a C$^\text{n}$NOT gate.

\section{\label{RD}Results and Discussion}
\subsection{Toffoli gate}

To implement the Toffoli gate using our proposed protocol, we first optimize the system parameters by analyzing the population dynamics. In our simulation, we use $^{87}$Rb atoms, and both target and control atoms are excited to $94S$ Rydberg state with the principal quantum number ($\mathcal{N}$) being 94 and the electronic orbital quantum number ($L$) being zero. The Rabi frequency of the laser that couples $\ket{1}$ and $\ket{r}$ of the control atom is chosen as $\Omega_r = \Omega_e$, so the duration of the first $\pi$ pulse is $\mathscr T_1 = \pi/\Omega_e$. For the Raman transition, we apply a smooth adiabatic pulse, 
\begin{equation}
    \Omega_e(t) = (\Omega_e/2)(1 - \cos(2\pi t/\mathscr T_2))
\end{equation}
with a pulse area of $3\Omega_e^2\mathscr T_2/(16\Delta)$. For a $\pi$ pulse, this yields 
\begin{equation}\label{ET2}
   \mathscr T_2 = 16\pi\Delta/(3\Omega_e^2).
\end{equation}
 To minimize decay of $\ket e$, we choose $\Delta \gg \gamma_e$, where $\gamma_e = 1/\tau_e$. Since the control atom is in $\ket{r}$ and decays at a rate $\gamma_r = 1/\tau_r$, the Raman pulse must be much faster, i.e. $\mathscr{T}_2 \ll \tau_r$. The time duration for the third $\pi$ pulse is 
\begin{equation}\label{ET1T2}
    \mathscr T_3 = \mathscr T_1 = \pi/\Omega_r
\end{equation}

\begin{figure}[!h]
\centering
     \includegraphics[width=6cm]{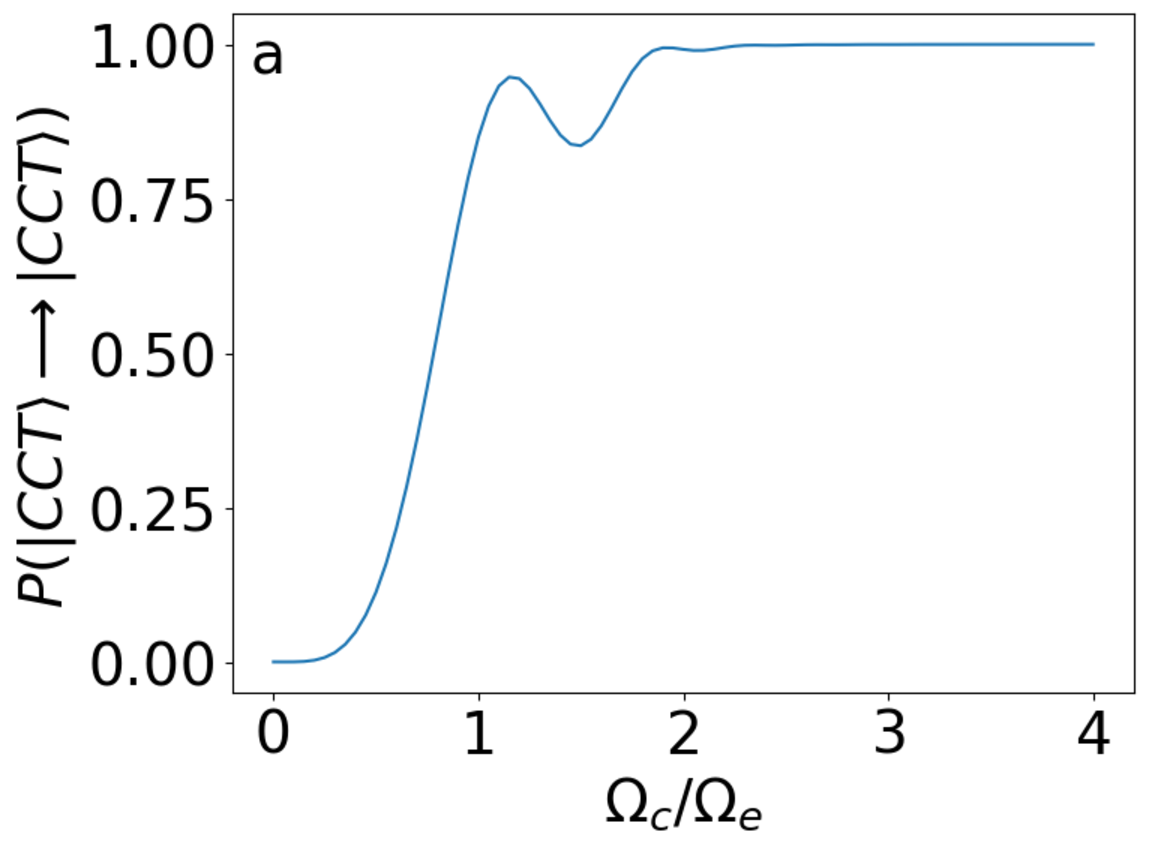}
     \includegraphics[width=6cm]{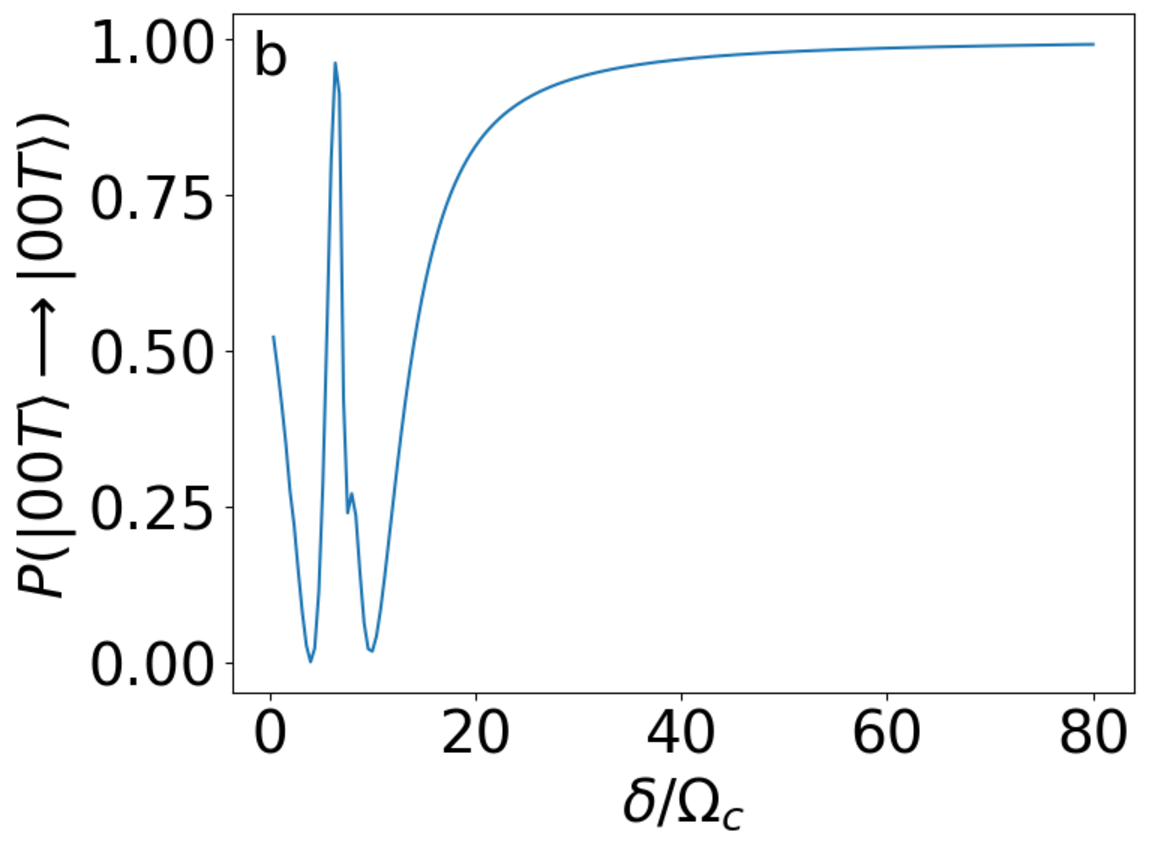}\\
\includegraphics[width=6cm]{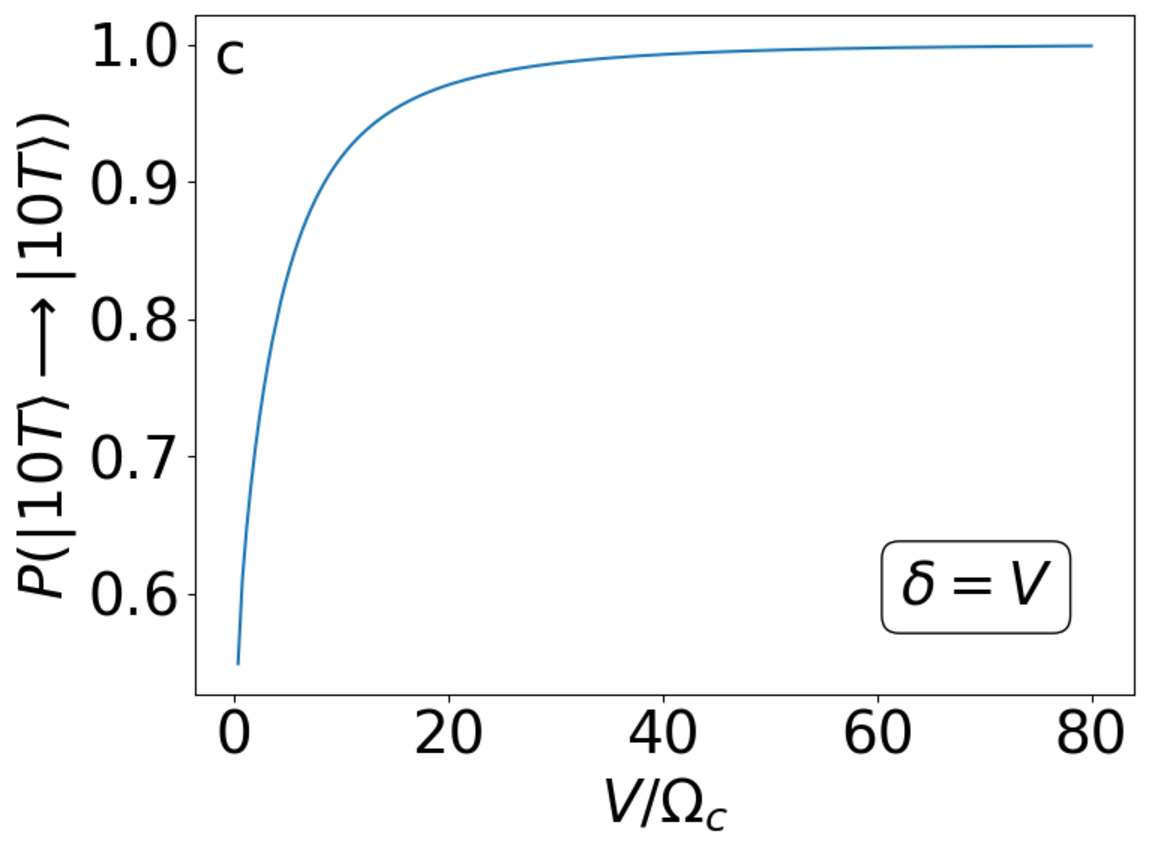}
     \includegraphics[width=6cm]{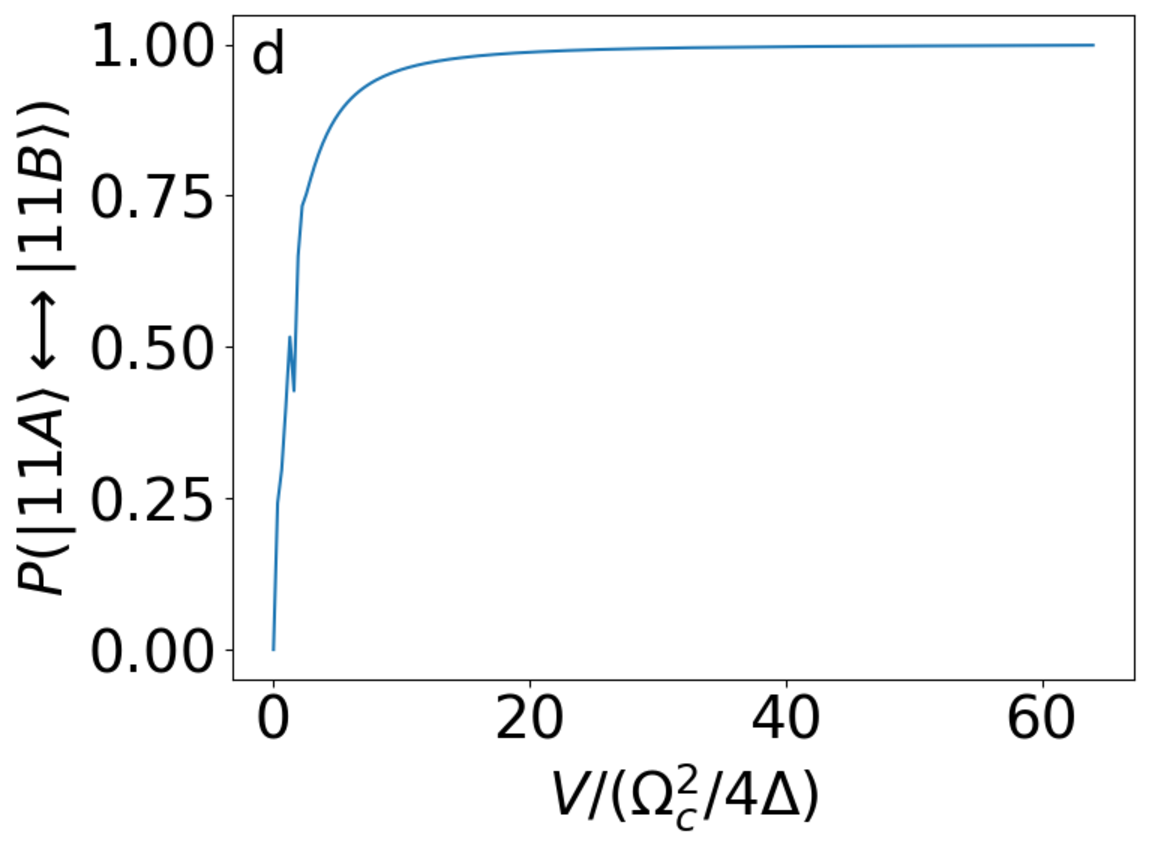}
    \caption{\label{F5} In the above diagrams $\ket C$ and $\ket{T}$ refer to the control and target qubits, respectively, where $C\in0,1$ and $T\in A,B$. (a) The probability of no transition of the target atom,  $\ket{CCT}\longrightarrow\ket{CCT}$ as a function of $\Omega_c/\Omega_e$ for $\delta\gg \Omega_c$ and $\ket{CC}\ne\ket{11}$. For $\Omega_c/\Omega_e>2$ the population remains unchanged, with more than $99\%$ probability. (b), (c) The probability of blocking the population of $\ket{T}$ for $\ket{CC}$ is $\ket {00}$, $\ket{10}$ or $\ket{01}$ respectively with $\Omega_c/\Omega_e=2.5$. For $\delta\gg \Omega_c$ we get the highest efficiency of blocking. (d) The transition probability $\ket{11A}\longleftrightarrow\ket{11B}$ as a function of interaction strength $V$ (in unit of $\Omega_c^2/4\Delta$)between the target and control atoms.   } 
 \end{figure}

Now, let us examine the population dynamics for various initial states. Fig.~\ref{F5}a illustrates the probability of $\ket{T}\longrightarrow\ket{T}$ ($\ket T$ is defined as the target qubit), that is the probability of no change in the target state, when both control atoms are initially in the state $\ket{0}$ or one is in $\ket{0}$ and the other in $\ket{1}$. For this plot, $\Omega_c/\Omega_e>2$, this figure shows that the population transfer is blocked with $>99\%$ probability for both initial states. Fig.~\ref{F5}b and \ref{F5}c show that for large $\delta $ the probability of $\ket{T}\longrightarrow\ket{T}$ gets maximum. By fixing $\Omega_c/\Omega_e =2.5$ and $\Delta/\Omega_e=10$ when both control atoms are initially in $\ket 1$, the probability of population transfer from $\ket{11A}$ to $\ket{11B}$ as a function of the control-target atom interaction strength $V $ is plotted and which shows that the probability increases for larger $V $ as Fig.~\ref{F5}d shows. 

 We take $\Omega_e/2\pi=44 \ \text{MHz}$ and the conversion between $V $ and $l$ is done using ($l=[C_6/V ]^{1/6}$) where $C_6$ is calculated using the formula defined as\cite{Singer2005, Loew2012}
\begin{equation}
    C_6=\mathcal{N}^{11}(11.97-0.8486\mathcal{N}+3.385\times10^{-3}\mathcal{N}^2) \ \text{au}.
\end{equation}
For an illustration, we consider the target-control atom distance $l=4 \ \mu$m for which the interaction strength can be calculated as $V=61\Omega_c$. We can neglect the interaction between the two control atoms $V_{cc} =0.96\Omega_c $, which are separated by $2l=8\ \mu$m. Substituting the values of $\Omega_e$ in the equations \ref{ET2} and \ref{ET1T2}, we calculate the total gate operation time $\mathscr T_{Tof}=(\mathscr T_1+\mathscr T_2+\mathscr T_3)<1 \ \mu$s ($T_1=T_3=11.3 \ \text{ns}, T_2=0.6 \ \mu\text{s}$).

  The average gate fidelity is computed using the trace-preserving, quantum-operator-based definition \cite{nielsen2002simple}:
\begin{equation}\label{Efid}
   \bar{F}(\hat{\mathcal{O}},\epsilon)
   = \frac{\sum_j \mathrm{tr}\bigl[\hat{\mathcal{O}}\,\hat{\mathcal{O}}_j^\dagger\,\hat{\mathcal{O}}^\dagger\,\epsilon(\hat{\mathcal{O}}_j)\bigr] + d^2}{d^2\,(d+1)},
\end{equation}
where $\hat{\mathcal{O}}_j$ denotes the set of all tensor products of n-qubit Pauli operators, for this case three-qubit operators 
$(\hat{I}\otimes\hat{I}\otimes\hat{I},\,\hat{I}\otimes\hat{I}\otimes\hat{\sigma}_x,\ldots,\hat{\sigma}_z\otimes\hat{\sigma}_z\otimes\hat{\sigma}_z)$,
$\hat{\mathcal{O}}$ is the ideal universal gate, for this case Toffoli gate, $\epsilon$ is the trace-preserving quantum operation obtained by our proposed quantum gate, and $d = 2^N$ is the dimension of the Hilbert space for an $N$-qubit system. 

In the presence of the spontaneous emission of Rydberg atoms, the evolution of the whole system can be governed by the master equation
\begin{equation}
    \dot{\hat{\rho}}=-i[\hat H, \hat \rho]+ \sum_{i=0}^1\mathscr{D}_1[\sigma_i]\hat{\rho}+\mathscr{D}_2[\alpha]\hat{\rho}+\sum_{j=A}^B\mathscr{D}_3[\beta_j]\hat{\rho},
\end{equation}
where $\mathscr {D}[\hat a]\hat \rho=a\hat \rho\hat a^\dagger-(\hat\rho\hat a^\dagger \hat a+\hat{a}^\dagger\hat a\hat \rho)/2$. The decay channels are defined as  $ \sigma_i=\sqrt{\gamma_r/2}\ket{i}\bra{r}$, $\alpha=\sqrt{\gamma_R}\ket{e}\bra{R} $ and $\beta_j=\sqrt{\gamma_e/2}\ket{j}\bra{e}$.
At $l=4 \ \mu$m and $\Delta=10\Omega_e$ we get the average gate fidelity  $\bar{F}_3\simeq96\%$. We use the lifetime of the state $\ket e$, same as mentioned in \cite{PhysRevLett.102.170502}, and the lifetime of $94S$ state of Rb$^{87}$ is taken as $100 \ \mu$s.

For the planner atomic configuration, the distance between all the atoms at the vertices of the equilateral triangle is $4\mu$m, so the interaction between the control atoms can not be neglected. Using the two-atom RAB mechanism, for the same atomic species and the same Rydberg state ($94S$) for the control atoms and the target atom, keeping all parameters same as before, we obtain the average gate fidelity  $\bar{F}_3'\simeq96\%$ at $0$ K, with the total gate time   $\mathscr T_{Tof}'=(2\mathscr T_1+\mathscr T_2+2\mathscr T_3)<1 \ \mu$s. 

\subsection{C$^3$NOT gate}
\begin{figure}[!h]
\centering
     \includegraphics[width=6cm]{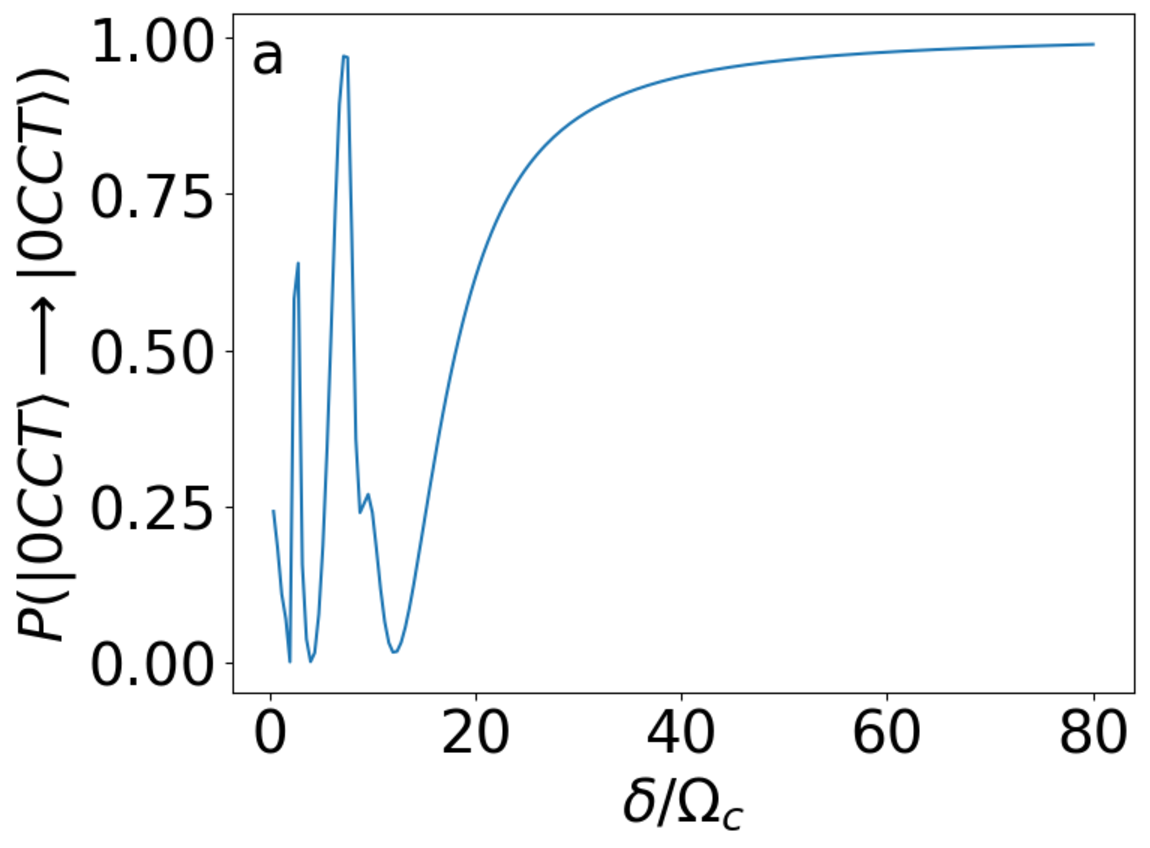}
     \includegraphics[width=6cm]{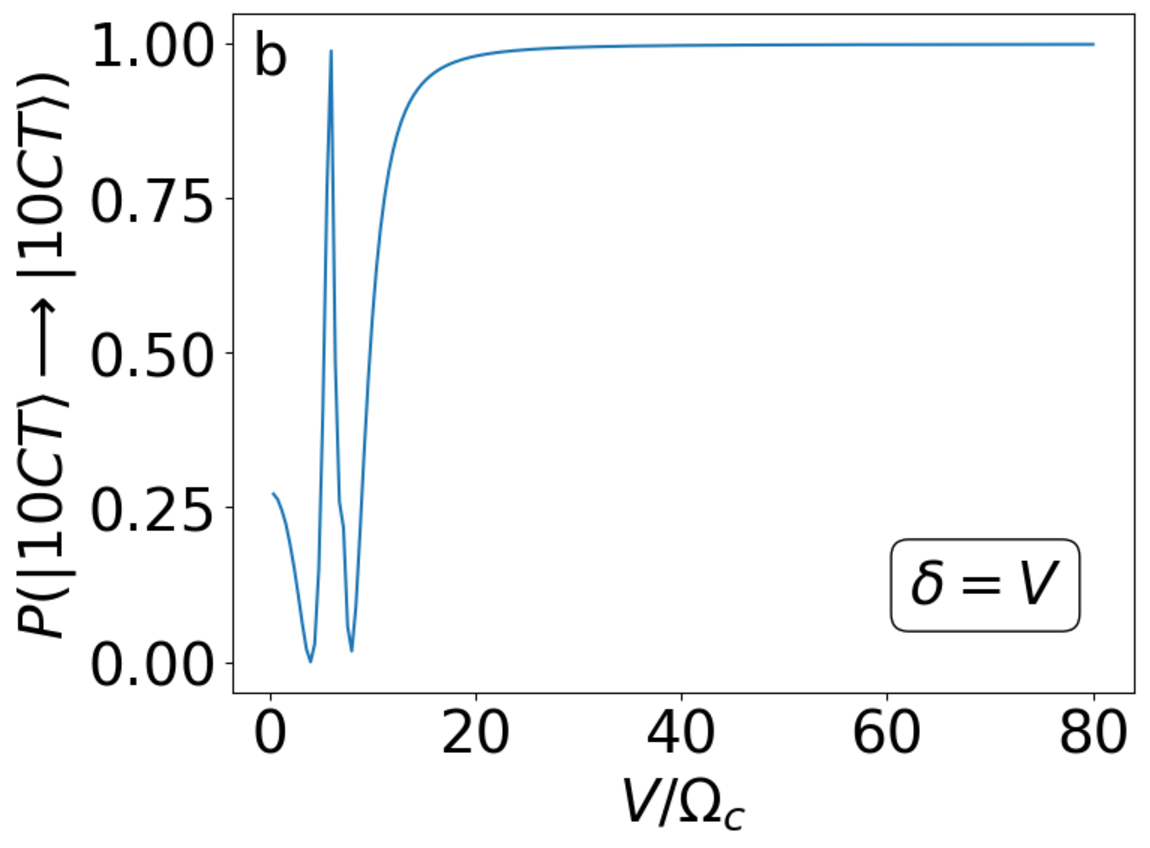}\\
\includegraphics[width=6cm]{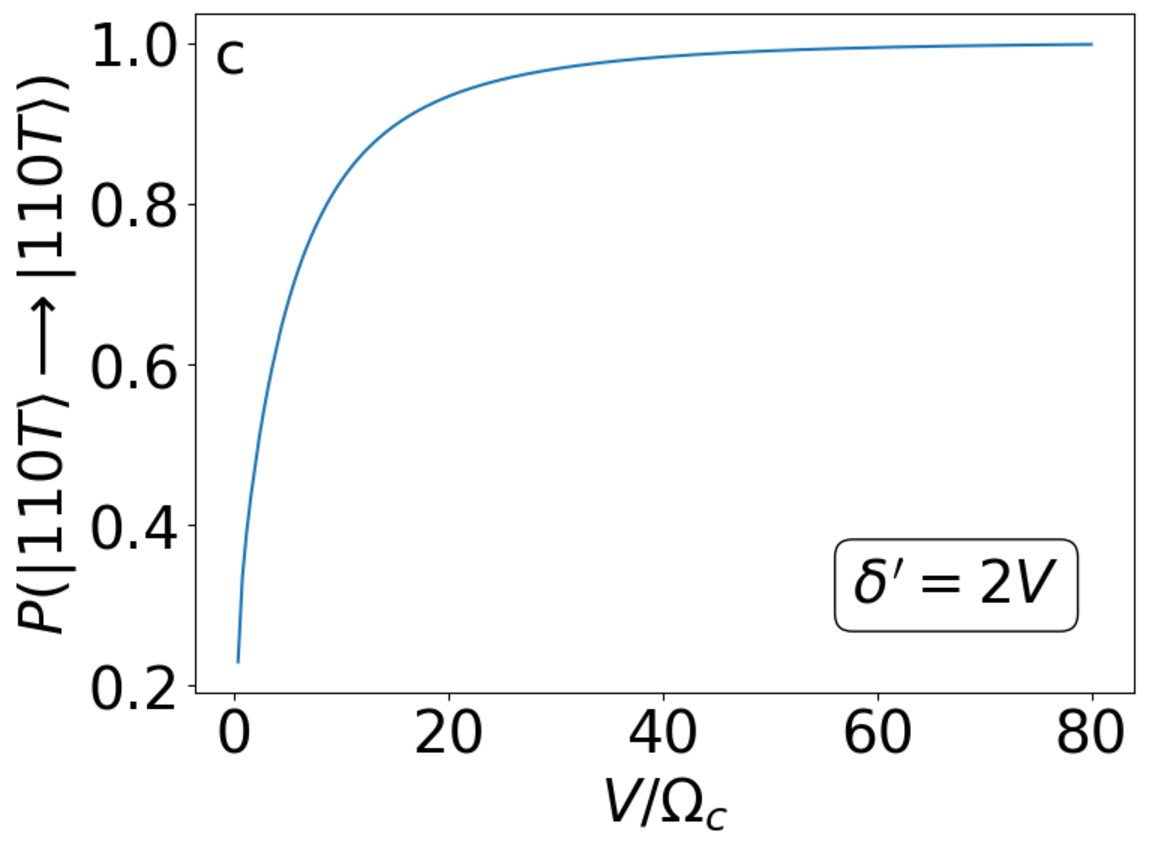}
     \includegraphics[width=6cm]{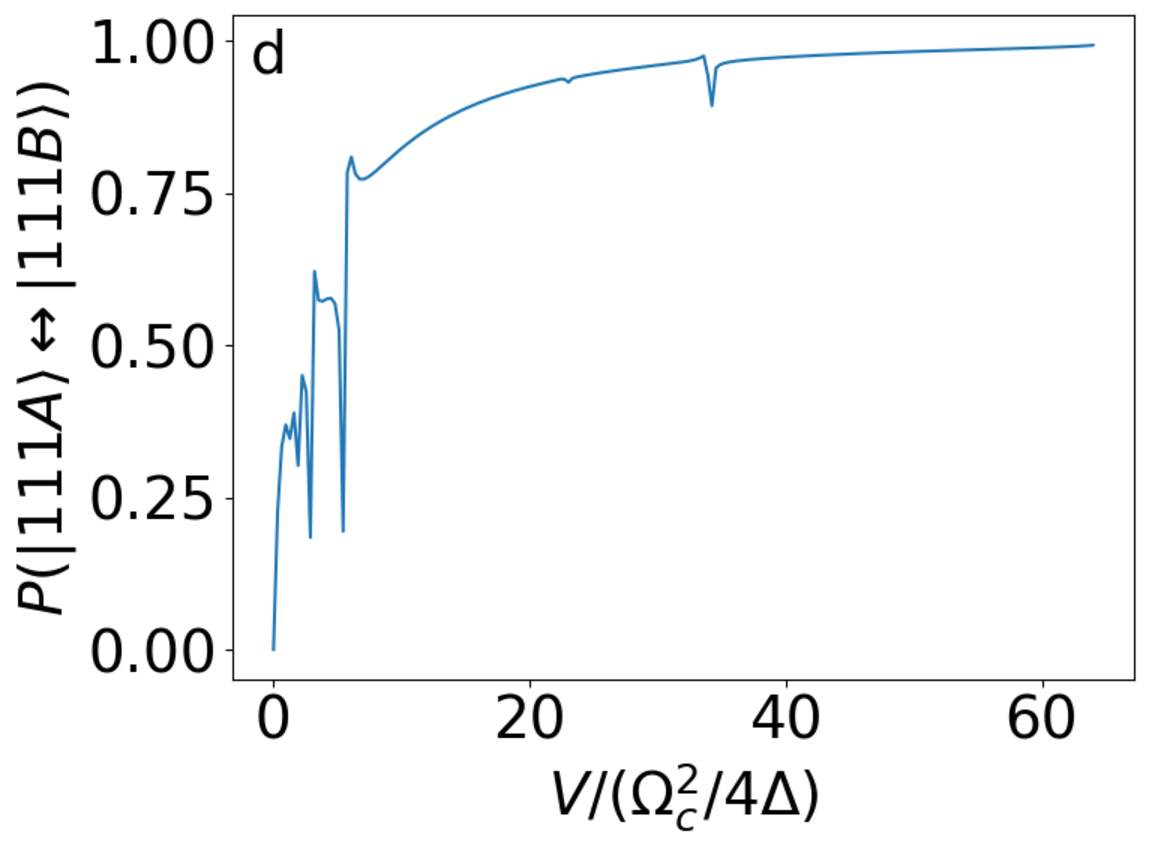}
    \caption{\label{F6} (a), (b), (c) The probability of $\ket{T}\longrightarrow\ket{T}$ for $\Omega_c/\Omega_e=2.5$ for three different initial conditions of the control atoms. For the detuning much larger than the $\Omega_c$, the probability of population blocking is maximum. (d) The probability of $\ket{111A}\longleftrightarrow\ket{111B}$ as a function of interaction strength (in the unit of $\Omega_c^2/4\Delta$) between the target and control atoms.} 
 \end{figure}

When the detuning of the lasers $\mathcal{L}_2$ and $\mathcal{L}_3$ are much larger than $\Omega_c$, we get the maximum blockage in population transfer $(\ket{T}\longrightarrow\ket{T})$ for $\Omega_c/\Omega_e>2$ similar to the previously explained Toffoli gate. Fixing $\Omega_c/\Omega_e=2.5$, we show the probability of population transfer blockage as a function of detuning or interaction strength for the three initial states of control atoms, $\ket{0CC},\ket{10C},\ket{110}$ in Fig.~\ref{F6}(a),(b),(c) respectively. Fig.~\ref{F6}(d)
shows the probability of population transfer from $\ket{A}\longleftrightarrow\ket{B}$ as a function of the control-target atom interaction potential when the control atoms are initially in $\ket{111}$ state. Now we take the distance between the target and control atoms as $4\mu$m, so the distance between the control atoms, which are situated at the vertices of the equilateral triangle, will be $6.93\mu$m. Applying the three-atom RAB mechanism and using the same physical conditions as mentioned above, we get the average gate fidelity $\bar{F}_4\simeq94\%$ using Eq.~(\ref{Efid}). The total gate time for the four-qubit C$^3$NOT gate is $\mathscr T_{{C}^3{NOT}}=(3\mathscr T_1+\mathscr T_2+3\mathscr T_3)<1 \ \mu$s.
 
 \section{Conclusions}
 In conclusion, we have demonstrated the Toffoli and C$^\text{n}$NOT ($n > 2$) gate operations using protocols based on Rydberg blockade, Rydberg antiblockade mechanism and dark state resonances. First, we have proposed a scheme for the Toffoli gate in a linear configuration of the three qubits and demonstrated the gate operation with 96\% fidelity.  Then,  to scale up the system, we have considered a triangular configuration of the qubits in a planar geometry. In this configuration, by making use of the two-atom RAB mechanism that allows populating the Rydberg states of both the control atoms at the same time,  we have demonstrated the Toffoli gate operation with about 96\% fidelity. Finally, we have constructed a C$^\text{3}$NOT gate using a three-atom RAB mechanism and generalized our protocol to realize C$^\text{n}$NOT gate with $n > 2$. We have numerically simulated the above-mentioned gates using realistic experimental parameters, including the spontaneous decay of the Rydberg atoms. Our protocols allow individual addressing of the qubits as they can be arranged with large inter-qubit spacing.

\bibliographystyle{apsrev4-2.bst} 
\bibliography{ref}  

\begin{thebibliography}{40}%
\makeatletter
\providecommand \@ifxundefined [1]{%
 \@ifx{#1\undefined}
}%
\providecommand \@ifnum [1]{%
 \ifnum #1\expandafter \@firstoftwo
 \else \expandafter \@secondoftwo
 \fi
}%
\providecommand \@ifx [1]{%
 \ifx #1\expandafter \@firstoftwo
 \else \expandafter \@secondoftwo
 \fi
}%
\providecommand \natexlab [1]{#1}%
\providecommand \enquote  [1]{``#1''}%
\providecommand \bibnamefont  [1]{#1}%
\providecommand \bibfnamefont [1]{#1}%
\providecommand \citenamefont [1]{#1}%
\providecommand \href@noop [0]{\@secondoftwo}%
\providecommand \href [0]{\begingroup \@sanitize@url \@href}%
\providecommand \@href[1]{\@@startlink{#1}\@@href}%
\providecommand \@@href[1]{\endgroup#1\@@endlink}%
\providecommand \@sanitize@url [0]{\catcode `\\12\catcode `\$12\catcode
  `\&12\catcode `\#12\catcode `\^12\catcode `\_12\catcode `\%12\relax}%
\providecommand \@@startlink[1]{}%
\providecommand \@@endlink[0]{}%
\providecommand \url  [0]{\begingroup\@sanitize@url \@url }%
\providecommand \@url [1]{\endgroup\@href {#1}{\urlprefix }}%
\providecommand \urlprefix  [0]{URL }%
\providecommand \Eprint [0]{\href }%
\providecommand \doibase [0]{https://doi.org/}%
\providecommand \selectlanguage [0]{\@gobble}%
\providecommand \bibinfo  [0]{\@secondoftwo}%
\providecommand \bibfield  [0]{\@secondoftwo}%
\providecommand \translation [1]{[#1]}%
\providecommand \BibitemOpen [0]{}%
\providecommand \bibitemStop [0]{}%
\providecommand \bibitemNoStop [0]{.\EOS\space}%
\providecommand \EOS [0]{\spacefactor3000\relax}%
\providecommand \BibitemShut  [1]{\csname bibitem#1\endcsname}%
\let\auto@bib@innerbib\@empty
\bibitem [{\citenamefont {Cirac}\ and\ \citenamefont
  {Zoller}(1995)}]{PhysRevLett.74.4091}%
  \BibitemOpen
  \bibfield  {author} {\bibinfo {author} {\bibfnamefont {J.~I.}\ \bibnamefont
  {Cirac}}\ and\ \bibinfo {author} {\bibfnamefont {P.}~\bibnamefont {Zoller}},\
  }\href {https://doi.org/10.1103/PhysRevLett.74.4091} {\bibfield  {journal}
  {\bibinfo  {journal} {Phys. Rev. Lett.}\ }\textbf {\bibinfo {volume} {74}},\
  \bibinfo {pages} {4091} (\bibinfo {year} {1995})}\BibitemShut {NoStop}%
\bibitem [{\citenamefont {Monroe}\ \emph {et~al.}(1995)\citenamefont {Monroe},
  \citenamefont {Meekhof}, \citenamefont {King}, \citenamefont {Itano},\ and\
  \citenamefont {Wineland}}]{Monroe1995Demonstration}%
  \BibitemOpen
  \bibfield  {author} {\bibinfo {author} {\bibfnamefont {C.}~\bibnamefont
  {Monroe}}, \bibinfo {author} {\bibfnamefont {D.~M.}\ \bibnamefont {Meekhof}},
  \bibinfo {author} {\bibfnamefont {B.~E.}\ \bibnamefont {King}}, \bibinfo
  {author} {\bibfnamefont {W.~M.}\ \bibnamefont {Itano}},\ and\ \bibinfo
  {author} {\bibfnamefont {D.~J.}\ \bibnamefont {Wineland}},\ }\href
  {https://doi.org/10.1103/PhysRevLett.75.4714} {\bibfield  {journal} {\bibinfo
   {journal} {Phys. Rev. Lett.}\ }\textbf {\bibinfo {volume} {75}},\ \bibinfo
  {pages} {4714} (\bibinfo {year} {1995})}\BibitemShut {NoStop}%
\bibitem [{\citenamefont {S\o{}rensen}\ and\ \citenamefont
  {M\o{}lmer}(2000)}]{PhysRevA.62.022311}%
  \BibitemOpen
  \bibfield  {author} {\bibinfo {author} {\bibfnamefont {A.}~\bibnamefont
  {S\o{}rensen}}\ and\ \bibinfo {author} {\bibfnamefont {K.}~\bibnamefont
  {M\o{}lmer}},\ }\href {https://doi.org/10.1103/PhysRevA.62.022311} {\bibfield
   {journal} {\bibinfo  {journal} {Phys. Rev. A}\ }\textbf {\bibinfo {volume}
  {62}},\ \bibinfo {pages} {022311} (\bibinfo {year} {2000})}\BibitemShut
  {NoStop}%
\bibitem [{\citenamefont {Sackett}\ \emph {et~al.}(2000)\citenamefont
  {Sackett}, \citenamefont {Kielpinski}, \citenamefont {King}, \citenamefont
  {Langer}, \citenamefont {Meyer}, \citenamefont {Myatt}, \citenamefont {Rowe},
  \citenamefont {Turchette}, \citenamefont {Itano}, \citenamefont {Wineland},\
  and\ \citenamefont {Monroe}}]{Sackett2000Experimental}%
  \BibitemOpen
  \bibfield  {author} {\bibinfo {author} {\bibfnamefont {C.~A.}\ \bibnamefont
  {Sackett}}, \bibinfo {author} {\bibfnamefont {D.}~\bibnamefont {Kielpinski}},
  \bibinfo {author} {\bibfnamefont {B.~E.}\ \bibnamefont {King}}, \bibinfo
  {author} {\bibfnamefont {C.}~\bibnamefont {Langer}}, \bibinfo {author}
  {\bibfnamefont {V.}~\bibnamefont {Meyer}}, \bibinfo {author} {\bibfnamefont
  {C.~J.}\ \bibnamefont {Myatt}}, \bibinfo {author} {\bibfnamefont
  {M.}~\bibnamefont {Rowe}}, \bibinfo {author} {\bibfnamefont {Q.~A.}\
  \bibnamefont {Turchette}}, \bibinfo {author} {\bibfnamefont {W.~M.}\
  \bibnamefont {Itano}}, \bibinfo {author} {\bibfnamefont {D.~J.}\ \bibnamefont
  {Wineland}},\ and\ \bibinfo {author} {\bibfnamefont {C.}~\bibnamefont
  {Monroe}},\ }\href {https://doi.org/10.1038/35005011} {\bibfield  {journal}
  {\bibinfo  {journal} {Nature}\ }\textbf {\bibinfo {volume} {404}},\ \bibinfo
  {pages} {256} (\bibinfo {year} {2000})}\BibitemShut {NoStop}%
\bibitem [{\citenamefont {Jaksch}\ \emph {et~al.}(2000)\citenamefont {Jaksch},
  \citenamefont {Cirac}, \citenamefont {Zoller}, \citenamefont {Rolston},
  \citenamefont {C\^ot\'e},\ and\ \citenamefont {Lukin}}]{PhysRevLett.85.2208}%
  \BibitemOpen
  \bibfield  {author} {\bibinfo {author} {\bibfnamefont {D.}~\bibnamefont
  {Jaksch}}, \bibinfo {author} {\bibfnamefont {J.~I.}\ \bibnamefont {Cirac}},
  \bibinfo {author} {\bibfnamefont {P.}~\bibnamefont {Zoller}}, \bibinfo
  {author} {\bibfnamefont {S.~L.}\ \bibnamefont {Rolston}}, \bibinfo {author}
  {\bibfnamefont {R.}~\bibnamefont {C\^ot\'e}},\ and\ \bibinfo {author}
  {\bibfnamefont {M.~D.}\ \bibnamefont {Lukin}},\ }\href
  {https://doi.org/10.1103/PhysRevLett.85.2208} {\bibfield  {journal} {\bibinfo
   {journal} {Phys. Rev. Lett.}\ }\textbf {\bibinfo {volume} {85}},\ \bibinfo
  {pages} {2208} (\bibinfo {year} {2000})}\BibitemShut {NoStop}%
\bibitem [{\citenamefont {Urban}\ \emph {et~al.}(2009)\citenamefont {Urban},
  \citenamefont {Johnson}, \citenamefont {Henage}, \citenamefont {Isenhower},
  \citenamefont {Yavuz}, \citenamefont {Walker},\ and\ \citenamefont
  {Saffman}}]{Urban2009Observation}%
  \BibitemOpen
  \bibfield  {author} {\bibinfo {author} {\bibfnamefont {E.}~\bibnamefont
  {Urban}}, \bibinfo {author} {\bibfnamefont {T.~A.}\ \bibnamefont {Johnson}},
  \bibinfo {author} {\bibfnamefont {T.}~\bibnamefont {Henage}}, \bibinfo
  {author} {\bibfnamefont {L.}~\bibnamefont {Isenhower}}, \bibinfo {author}
  {\bibfnamefont {D.~D.}\ \bibnamefont {Yavuz}}, \bibinfo {author}
  {\bibfnamefont {T.~G.}\ \bibnamefont {Walker}},\ and\ \bibinfo {author}
  {\bibfnamefont {M.}~\bibnamefont {Saffman}},\ }\href
  {https://doi.org/10.1038/nphys1178} {\bibfield  {journal} {\bibinfo
  {journal} {Nature Physics}\ }\textbf {\bibinfo {volume} {5}},\ \bibinfo
  {pages} {110} (\bibinfo {year} {2009})}\BibitemShut {NoStop}%
\bibitem [{\citenamefont {Isenhower}\ \emph {et~al.}(2010)\citenamefont
  {Isenhower}, \citenamefont {Urban}, \citenamefont {Zhang}, \citenamefont
  {Gill}, \citenamefont {Henage}, \citenamefont {Johnson}, \citenamefont
  {Walker},\ and\ \citenamefont {Saffman}}]{PhysRevLett.104.010503}%
  \BibitemOpen
  \bibfield  {author} {\bibinfo {author} {\bibfnamefont {L.}~\bibnamefont
  {Isenhower}}, \bibinfo {author} {\bibfnamefont {E.}~\bibnamefont {Urban}},
  \bibinfo {author} {\bibfnamefont {X.~L.}\ \bibnamefont {Zhang}}, \bibinfo
  {author} {\bibfnamefont {A.~T.}\ \bibnamefont {Gill}}, \bibinfo {author}
  {\bibfnamefont {T.}~\bibnamefont {Henage}}, \bibinfo {author} {\bibfnamefont
  {T.~A.}\ \bibnamefont {Johnson}}, \bibinfo {author} {\bibfnamefont {T.~G.}\
  \bibnamefont {Walker}},\ and\ \bibinfo {author} {\bibfnamefont
  {M.}~\bibnamefont {Saffman}},\ }\href
  {https://doi.org/10.1103/PhysRevLett.104.010503} {\bibfield  {journal}
  {\bibinfo  {journal} {Phys. Rev. Lett.}\ }\textbf {\bibinfo {volume} {104}},\
  \bibinfo {pages} {010503} (\bibinfo {year} {2010})}\BibitemShut {NoStop}%
\bibitem [{\citenamefont {Maller}\ \emph {et~al.}(2015)\citenamefont {Maller},
  \citenamefont {Lichtman}, \citenamefont {Xia}, \citenamefont {Sun},
  \citenamefont {Piotrowicz}, \citenamefont {Carr}, \citenamefont {Isenhower},\
  and\ \citenamefont {Saffman}}]{PhysRevA.92.022336}%
  \BibitemOpen
  \bibfield  {author} {\bibinfo {author} {\bibfnamefont {K.~M.}\ \bibnamefont
  {Maller}}, \bibinfo {author} {\bibfnamefont {M.~T.}\ \bibnamefont
  {Lichtman}}, \bibinfo {author} {\bibfnamefont {T.}~\bibnamefont {Xia}},
  \bibinfo {author} {\bibfnamefont {Y.}~\bibnamefont {Sun}}, \bibinfo {author}
  {\bibfnamefont {M.~J.}\ \bibnamefont {Piotrowicz}}, \bibinfo {author}
  {\bibfnamefont {A.~W.}\ \bibnamefont {Carr}}, \bibinfo {author}
  {\bibfnamefont {L.}~\bibnamefont {Isenhower}},\ and\ \bibinfo {author}
  {\bibfnamefont {M.}~\bibnamefont {Saffman}},\ }\href
  {https://doi.org/10.1103/PhysRevA.92.022336} {\bibfield  {journal} {\bibinfo
  {journal} {Phys. Rev. A}\ }\textbf {\bibinfo {volume} {92}},\ \bibinfo
  {pages} {022336} (\bibinfo {year} {2015})}\BibitemShut {NoStop}%
\bibitem [{\citenamefont {Theis}\ \emph {et~al.}(2016)\citenamefont {Theis},
  \citenamefont {Motzoi}, \citenamefont {Wilhelm},\ and\ \citenamefont
  {Saffman}}]{PhysRevA.94.032306}%
  \BibitemOpen
  \bibfield  {author} {\bibinfo {author} {\bibfnamefont {L.~S.}\ \bibnamefont
  {Theis}}, \bibinfo {author} {\bibfnamefont {F.}~\bibnamefont {Motzoi}},
  \bibinfo {author} {\bibfnamefont {F.~K.}\ \bibnamefont {Wilhelm}},\ and\
  \bibinfo {author} {\bibfnamefont {M.}~\bibnamefont {Saffman}},\ }\href
  {https://doi.org/10.1103/PhysRevA.94.032306} {\bibfield  {journal} {\bibinfo
  {journal} {Phys. Rev. A}\ }\textbf {\bibinfo {volume} {94}},\ \bibinfo
  {pages} {032306} (\bibinfo {year} {2016})}\BibitemShut {NoStop}%
\bibitem [{\citenamefont {Shi}(2017)}]{Shi2017RydbergGatesFree}%
  \BibitemOpen
  \bibfield  {author} {\bibinfo {author} {\bibfnamefont {X.}~\bibnamefont
  {Shi}},\ }\href {https://doi.org/10.1103/PhysRevApplied.7.064017} {\bibfield
  {journal} {\bibinfo  {journal} {Phys. Rev. Applied}\ }\textbf {\bibinfo
  {volume} {7}},\ \bibinfo {pages} {064017} (\bibinfo {year} {2017})},\ \Eprint
  {https://arxiv.org/abs/1611.00750} {arXiv:1611.00750 [quant-ph]} \BibitemShut
  {NoStop}%
\bibitem [{\citenamefont {Shi}\ and\ \citenamefont
  {Lu}(2021)}]{PhysRevA.104.012615}%
  \BibitemOpen
  \bibfield  {author} {\bibinfo {author} {\bibfnamefont {X.-F.}\ \bibnamefont
  {Shi}}\ and\ \bibinfo {author} {\bibfnamefont {Y.}~\bibnamefont {Lu}},\
  }\href {https://doi.org/10.1103/PhysRevA.104.012615} {\bibfield  {journal}
  {\bibinfo  {journal} {Phys. Rev. A}\ }\textbf {\bibinfo {volume} {104}},\
  \bibinfo {pages} {012615} (\bibinfo {year} {2021})}\BibitemShut {NoStop}%
\bibitem [{\citenamefont {Levine}\ \emph {et~al.}(2019)\citenamefont {Levine},
  \citenamefont {Keesling}, \citenamefont {Omran}, \citenamefont {Bernien},
  \citenamefont {Schwartz}, \citenamefont {Zibrov}, \citenamefont {Endres},
  \citenamefont {Greiner}, \citenamefont {Vuleti{\'c}},\ and\ \citenamefont
  {Lukin}}]{levine2019parallel}%
  \BibitemOpen
  \bibfield  {author} {\bibinfo {author} {\bibfnamefont {H.}~\bibnamefont
  {Levine}}, \bibinfo {author} {\bibfnamefont {A.}~\bibnamefont {Keesling}},
  \bibinfo {author} {\bibfnamefont {A.}~\bibnamefont {Omran}}, \bibinfo
  {author} {\bibfnamefont {H.}~\bibnamefont {Bernien}}, \bibinfo {author}
  {\bibfnamefont {S.}~\bibnamefont {Schwartz}}, \bibinfo {author}
  {\bibfnamefont {A.~S.}\ \bibnamefont {Zibrov}}, \bibinfo {author}
  {\bibfnamefont {M.}~\bibnamefont {Endres}}, \bibinfo {author} {\bibfnamefont
  {M.}~\bibnamefont {Greiner}}, \bibinfo {author} {\bibfnamefont
  {V.}~\bibnamefont {Vuleti{\'c}}},\ and\ \bibinfo {author} {\bibfnamefont
  {M.~D.}\ \bibnamefont {Lukin}},\ }\href
  {https://doi.org/10.1103/PhysRevLett.123.170503} {\bibfield  {journal}
  {\bibinfo  {journal} {Phys. Rev. Lett.}\ }\textbf {\bibinfo {volume} {123}},\
  \bibinfo {pages} {170503} (\bibinfo {year} {2019})}\BibitemShut {NoStop}%
\bibitem [{\citenamefont {Maller}\ \emph {et~al.}(2017)\citenamefont {Maller},
  \citenamefont {Lichtman}, \citenamefont {Xia}, \citenamefont {Sun},
  \citenamefont {Piotrowicz}, \citenamefont {Carr}, \citenamefont {Isenhower},\
  and\ \citenamefont {Saffman}}]{maller2017high}%
  \BibitemOpen
  \bibfield  {author} {\bibinfo {author} {\bibfnamefont {K.~M.}\ \bibnamefont
  {Maller}}, \bibinfo {author} {\bibfnamefont {M.~T.}\ \bibnamefont
  {Lichtman}}, \bibinfo {author} {\bibfnamefont {T.}~\bibnamefont {Xia}},
  \bibinfo {author} {\bibfnamefont {Y.-T.}\ \bibnamefont {Sun}}, \bibinfo
  {author} {\bibfnamefont {M.~J.}\ \bibnamefont {Piotrowicz}}, \bibinfo
  {author} {\bibfnamefont {A.~W.}\ \bibnamefont {Carr}}, \bibinfo {author}
  {\bibfnamefont {L.}~\bibnamefont {Isenhower}},\ and\ \bibinfo {author}
  {\bibfnamefont {M.}~\bibnamefont {Saffman}},\ }\href
  {https://doi.org/10.1103/PhysRevA.96.042306} {\bibfield  {journal} {\bibinfo
  {journal} {Phys. Rev. A}\ }\textbf {\bibinfo {volume} {96}},\ \bibinfo
  {pages} {042306} (\bibinfo {year} {2017})}\BibitemShut {NoStop}%
\bibitem [{\citenamefont {M{\"u}ller}\ \emph {et~al.}(2009)\citenamefont
  {M{\"u}ller}, \citenamefont {Lesanovsky}, \citenamefont {Weimer},
  \citenamefont {B{\"u}chler},\ and\ \citenamefont
  {Zoller}}]{PhysRevLett.102.170502}%
  \BibitemOpen
  \bibfield  {author} {\bibinfo {author} {\bibfnamefont {M.}~\bibnamefont
  {M{\"u}ller}}, \bibinfo {author} {\bibfnamefont {I.}~\bibnamefont
  {Lesanovsky}}, \bibinfo {author} {\bibfnamefont {H.}~\bibnamefont {Weimer}},
  \bibinfo {author} {\bibfnamefont {H.~P.}\ \bibnamefont {B{\"u}chler}},\ and\
  \bibinfo {author} {\bibfnamefont {P.}~\bibnamefont {Zoller}},\ }\href
  {https://doi.org/10.1103/PhysRevLett.102.170502} {\bibfield  {journal}
  {\bibinfo  {journal} {Phys. Rev. Lett.}\ }\textbf {\bibinfo {volume} {102}},\
  \bibinfo {pages} {170502} (\bibinfo {year} {2009})}\BibitemShut {NoStop}%
\bibitem [{\citenamefont {Petrosyan}\ \emph {et~al.}(2017)\citenamefont
  {Petrosyan}, \citenamefont {Motzoi}, \citenamefont {Saffman},\ and\
  \citenamefont {M{\o}lmer}}]{Petrosyan2017}%
  \BibitemOpen
  \bibfield  {author} {\bibinfo {author} {\bibfnamefont {D.}~\bibnamefont
  {Petrosyan}}, \bibinfo {author} {\bibfnamefont {F.}~\bibnamefont {Motzoi}},
  \bibinfo {author} {\bibfnamefont {M.}~\bibnamefont {Saffman}},\ and\ \bibinfo
  {author} {\bibfnamefont {K.}~\bibnamefont {M{\o}lmer}},\ }\href
  {https://doi.org/10.1103/PhysRevA.96.042306} {\bibfield  {journal} {\bibinfo
  {journal} {Phys. Rev. A}\ }\textbf {\bibinfo {volume} {96}},\ \bibinfo
  {pages} {042306} (\bibinfo {year} {2017})}\BibitemShut {NoStop}%
\bibitem [{\citenamefont {Liu}\ \emph {et~al.}(2022)\citenamefont {Liu},
  \citenamefont {Shen}, \citenamefont {Zheng}, \citenamefont {Kang},
  \citenamefont {Shi}, \citenamefont {Song},\ and\ \citenamefont
  {Xia}}]{Liu2022}%
  \BibitemOpen
  \bibfield  {author} {\bibinfo {author} {\bibfnamefont {S.}~\bibnamefont
  {Liu}}, \bibinfo {author} {\bibfnamefont {J.-H.}\ \bibnamefont {Shen}},
  \bibinfo {author} {\bibfnamefont {R.-H.}\ \bibnamefont {Zheng}}, \bibinfo
  {author} {\bibfnamefont {Y.-H.}\ \bibnamefont {Kang}}, \bibinfo {author}
  {\bibfnamefont {Z.-C.}\ \bibnamefont {Shi}}, \bibinfo {author} {\bibfnamefont
  {J.}~\bibnamefont {Song}},\ and\ \bibinfo {author} {\bibfnamefont
  {Y.}~\bibnamefont {Xia}},\ }\href {https://doi.org/10.1007/s11467-021-1108-3}
  {\bibfield  {journal} {\bibinfo  {journal} {Front. Phys.}\ }\textbf {\bibinfo
  {volume} {17}},\ \bibinfo {pages} {21502} (\bibinfo {year}
  {2022})}\BibitemShut {NoStop}%
\bibitem [{\citenamefont {Shen}\ \emph {et~al.}(2019)\citenamefont {Shen},
  \citenamefont {Wu}, \citenamefont {Su},\ and\ \citenamefont
  {Liang}}]{Shen2019}%
  \BibitemOpen
  \bibfield  {author} {\bibinfo {author} {\bibfnamefont {C.-P.}\ \bibnamefont
  {Shen}}, \bibinfo {author} {\bibfnamefont {J.-L.}\ \bibnamefont {Wu}},
  \bibinfo {author} {\bibfnamefont {S.-L.}\ \bibnamefont {Su}},\ and\ \bibinfo
  {author} {\bibfnamefont {E.}~\bibnamefont {Liang}},\ }\href
  {https://doi.org/10.1364/OL.44.002036} {\bibfield  {journal} {\bibinfo
  {journal} {Opt. Lett.}\ }\textbf {\bibinfo {volume} {44}},\ \bibinfo {pages}
  {2036} (\bibinfo {year} {2019})}\BibitemShut {NoStop}%
\bibitem [{\citenamefont {Guo}\ \emph {et~al.}(2020)\citenamefont {Guo},
  \citenamefont {Yan}, \citenamefont {Zhang}, \citenamefont {Su},\ and\
  \citenamefont {Li}}]{Guo2020}%
  \BibitemOpen
  \bibfield  {author} {\bibinfo {author} {\bibfnamefont {C.-Y.}\ \bibnamefont
  {Guo}}, \bibinfo {author} {\bibfnamefont {L.-L.}\ \bibnamefont {Yan}},
  \bibinfo {author} {\bibfnamefont {S.}~\bibnamefont {Zhang}}, \bibinfo
  {author} {\bibfnamefont {S.-L.}\ \bibnamefont {Su}},\ and\ \bibinfo {author}
  {\bibfnamefont {W.}~\bibnamefont {Li}},\ }\href
  {https://doi.org/10.1103/PhysRevA.102.042607} {\bibfield  {journal} {\bibinfo
   {journal} {Phys. Rev. A}\ }\textbf {\bibinfo {volume} {102}},\ \bibinfo
  {pages} {042607} (\bibinfo {year} {2020})}\BibitemShut {NoStop}%
\bibitem [{\citenamefont {Xue}\ \emph {et~al.}(2024)\citenamefont {Xue},
  \citenamefont {Xu}, \citenamefont {Li},\ and\ \citenamefont {Li}}]{Xue2024}%
  \BibitemOpen
  \bibfield  {author} {\bibinfo {author} {\bibfnamefont {M.}~\bibnamefont
  {Xue}}, \bibinfo {author} {\bibfnamefont {S.}~\bibnamefont {Xu}}, \bibinfo
  {author} {\bibfnamefont {X.}~\bibnamefont {Li}},\ and\ \bibinfo {author}
  {\bibfnamefont {X.}~\bibnamefont {Li}},\ }\href
  {https://doi.org/10.1103/PhysRevA.110.032619} {\bibfield  {journal} {\bibinfo
   {journal} {Phys. Rev. A}\ }\textbf {\bibinfo {volume} {110}},\ \bibinfo
  {pages} {032619} (\bibinfo {year} {2024})}\BibitemShut {NoStop}%
\bibitem [{\citenamefont {Zhao}\ \emph {et~al.}(2018)\citenamefont {Zhao},
  \citenamefont {Wu}, \citenamefont {Xing}, \citenamefont {Xu},\ and\
  \citenamefont {Tong}}]{Zhao2018}%
  \BibitemOpen
  \bibfield  {author} {\bibinfo {author} {\bibfnamefont {P.~Z.}\ \bibnamefont
  {Zhao}}, \bibinfo {author} {\bibfnamefont {X.}~\bibnamefont {Wu}}, \bibinfo
  {author} {\bibfnamefont {T.~H.}\ \bibnamefont {Xing}}, \bibinfo {author}
  {\bibfnamefont {G.~F.}\ \bibnamefont {Xu}},\ and\ \bibinfo {author}
  {\bibfnamefont {D.~M.}\ \bibnamefont {Tong}},\ }\href
  {https://doi.org/10.1103/PhysRevA.98.032313} {\bibfield  {journal} {\bibinfo
  {journal} {Phys. Rev. A}\ }\textbf {\bibinfo {volume} {98}},\ \bibinfo
  {pages} {032313} (\bibinfo {year} {2018})}\BibitemShut {NoStop}%
\bibitem [{\citenamefont {Jin}\ and\ \citenamefont {Jing}(2024)}]{Jin2024}%
  \BibitemOpen
  \bibfield  {author} {\bibinfo {author} {\bibfnamefont {Z.-y.}\ \bibnamefont
  {Jin}}\ and\ \bibinfo {author} {\bibfnamefont {J.}~\bibnamefont {Jing}},\
  }\href {https://doi.org/10.1103/PhysRevA.109.012619} {\bibfield  {journal}
  {\bibinfo  {journal} {Phys. Rev. A}\ }\textbf {\bibinfo {volume} {109}},\
  \bibinfo {pages} {012619} (\bibinfo {year} {2024})}\BibitemShut {NoStop}%
\bibitem [{\citenamefont {Su}\ \emph {et~al.}(2023)\citenamefont {Su},
  \citenamefont {Sun}, \citenamefont {Liu}, \citenamefont {Yan}, \citenamefont
  {Yung}, \citenamefont {Li},\ and\ \citenamefont {Feng}}]{Su2023RabiBlockade}%
  \BibitemOpen
  \bibfield  {author} {\bibinfo {author} {\bibfnamefont {S.-L.}\ \bibnamefont
  {Su}}, \bibinfo {author} {\bibfnamefont {L.-N.}\ \bibnamefont {Sun}},
  \bibinfo {author} {\bibfnamefont {B.-J.}\ \bibnamefont {Liu}}, \bibinfo
  {author} {\bibfnamefont {L.-L.}\ \bibnamefont {Yan}}, \bibinfo {author}
  {\bibfnamefont {M.-H.}\ \bibnamefont {Yung}}, \bibinfo {author}
  {\bibfnamefont {W.}~\bibnamefont {Li}},\ and\ \bibinfo {author}
  {\bibfnamefont {M.}~\bibnamefont {Feng}},\ }\href
  {https://doi.org/10.1103/PhysRevApplied.19.044007} {\bibfield  {journal}
  {\bibinfo  {journal} {Phys. Rev. Applied}\ }\textbf {\bibinfo {volume}
  {19}},\ \bibinfo {pages} {044007} (\bibinfo {year} {2023})}\BibitemShut
  {NoStop}%
\bibitem [{\citenamefont {Mudli}\ \emph {et~al.}(2024)\citenamefont {Mudli},
  \citenamefont {Mal}, \citenamefont {Rej}, \citenamefont {Dey},\ and\
  \citenamefont {Deb}}]{PhysRevA.110.062618}%
  \BibitemOpen
  \bibfield  {author} {\bibinfo {author} {\bibfnamefont {S.}~\bibnamefont
  {Mudli}}, \bibinfo {author} {\bibfnamefont {S.}~\bibnamefont {Mal}}, \bibinfo
  {author} {\bibfnamefont {S.~S.}\ \bibnamefont {Rej}}, \bibinfo {author}
  {\bibfnamefont {A.}~\bibnamefont {Dey}},\ and\ \bibinfo {author}
  {\bibfnamefont {B.}~\bibnamefont {Deb}},\ }\href
  {https://doi.org/10.1103/PhysRevA.110.062618} {\bibfield  {journal} {\bibinfo
   {journal} {Phys. Rev. A}\ }\textbf {\bibinfo {volume} {110}},\ \bibinfo
  {pages} {062618} (\bibinfo {year} {2024})}\BibitemShut {NoStop}%
\bibitem [{\citenamefont {Su}\ \emph {et~al.}(2017)\citenamefont {Su},
  \citenamefont {Gao}, \citenamefont {Liang},\ and\ \citenamefont
  {Zhang}}]{PhysRevA.95.022319}%
  \BibitemOpen
  \bibfield  {author} {\bibinfo {author} {\bibfnamefont {S.-L.}\ \bibnamefont
  {Su}}, \bibinfo {author} {\bibfnamefont {Y.}~\bibnamefont {Gao}}, \bibinfo
  {author} {\bibfnamefont {E.}~\bibnamefont {Liang}},\ and\ \bibinfo {author}
  {\bibfnamefont {S.}~\bibnamefont {Zhang}},\ }\href
  {https://doi.org/10.1103/PhysRevA.95.022319} {\bibfield  {journal} {\bibinfo
  {journal} {Phys. Rev. A}\ }\textbf {\bibinfo {volume} {95}},\ \bibinfo
  {pages} {022319} (\bibinfo {year} {2017})}\BibitemShut {NoStop}%
\bibitem [{\citenamefont {Wu}\ \emph {et~al.}(2021{\natexlab{a}})\citenamefont
  {Wu}, \citenamefont {Wang}, \citenamefont {Han}, \citenamefont {Feng},
  \citenamefont {Su}, \citenamefont {Xia}, \citenamefont {Jiang},\ and\
  \citenamefont {Song}}]{Wu2021AntiblockadeSWAP}%
  \BibitemOpen
  \bibfield  {author} {\bibinfo {author} {\bibfnamefont {J.}~\bibnamefont
  {Wu}}, \bibinfo {author} {\bibfnamefont {Y.}~\bibnamefont {Wang}}, \bibinfo
  {author} {\bibfnamefont {J.}~\bibnamefont {Han}}, \bibinfo {author}
  {\bibfnamefont {Y.}~\bibnamefont {Feng}}, \bibinfo {author} {\bibfnamefont
  {S.}~\bibnamefont {Su}}, \bibinfo {author} {\bibfnamefont {Y.}~\bibnamefont
  {Xia}}, \bibinfo {author} {\bibfnamefont {Y.}~\bibnamefont {Jiang}},\ and\
  \bibinfo {author} {\bibfnamefont {J.}~\bibnamefont {Song}},\ }\href
  {https://opg.optica.org/prj/abstract.cfm?uri=prj-9-5-814} {\bibfield
  {journal} {\bibinfo  {journal} {Photon.\ Res.}\ }\textbf {\bibinfo {volume}
  {9}},\ \bibinfo {pages} {814} (\bibinfo {year} {2021}{\natexlab{a}})},\
  \bibinfo {note} {© 2021 Optical Society of America}\BibitemShut {NoStop}%
\bibitem [{\citenamefont {Wu}\ \emph {et~al.}(2021{\natexlab{b}})\citenamefont
  {Wu}, \citenamefont {Wang}, \citenamefont {Han}, \citenamefont {Su},
  \citenamefont {Xia}, \citenamefont {Jiang},\ and\ \citenamefont
  {Song}}]{Wu2021ResilientRydberg}%
  \BibitemOpen
  \bibfield  {author} {\bibinfo {author} {\bibfnamefont {J.}~\bibnamefont
  {Wu}}, \bibinfo {author} {\bibfnamefont {Y.}~\bibnamefont {Wang}}, \bibinfo
  {author} {\bibfnamefont {J.}~\bibnamefont {Han}}, \bibinfo {author}
  {\bibfnamefont {S.}~\bibnamefont {Su}}, \bibinfo {author} {\bibfnamefont
  {Y.}~\bibnamefont {Xia}}, \bibinfo {author} {\bibfnamefont {Y.}~\bibnamefont
  {Jiang}},\ and\ \bibinfo {author} {\bibfnamefont {J.}~\bibnamefont {Song}},\
  }\href {https://doi.org/10.1103/PhysRevA.103.012601} {\bibfield  {journal}
  {\bibinfo  {journal} {Phys. Rev. A}\ }\textbf {\bibinfo {volume} {103}},\
  \bibinfo {pages} {012601} (\bibinfo {year} {2021}{\natexlab{b}})},\ \Eprint
  {https://arxiv.org/abs/2101.02328} {arXiv:2101.02328 [quant-ph]} \BibitemShut
  {NoStop}%
\bibitem [{\citenamefont {Li}\ \emph {et~al.}(2023)\citenamefont {Li},
  \citenamefont {Wu}, \citenamefont {Su},\ and\ \citenamefont
  {Qian}}]{Li2023HighToleranceAntiblockade}%
  \BibitemOpen
  \bibfield  {author} {\bibinfo {author} {\bibfnamefont {W.}~\bibnamefont
  {Li}}, \bibinfo {author} {\bibfnamefont {J.}~\bibnamefont {Wu}}, \bibinfo
  {author} {\bibfnamefont {S.}~\bibnamefont {Su}},\ and\ \bibinfo {author}
  {\bibfnamefont {J.}~\bibnamefont {Qian}},\ }\href
  {https://arxiv.org/abs/2309.06013} {\bibinfo {title} {High‐tolerance
  antiblockade swap gates using optimal pulse drivings}},\ \bibinfo
  {howpublished} {arXiv:2309.06013 [quant-ph]} (\bibinfo {year}
  {2023})\BibitemShut {NoStop}%
\bibitem [{\citenamefont {Nielsen}\ and\ \citenamefont
  {Chuang}(2000)}]{nielsen2000quantum}%
  \BibitemOpen
  \bibfield  {author} {\bibinfo {author} {\bibfnamefont {M.~A.}\ \bibnamefont
  {Nielsen}}\ and\ \bibinfo {author} {\bibfnamefont {I.~L.}\ \bibnamefont
  {Chuang}},\ }\href {https://doi.org/10.1017/CBO9780511976667} {\emph
  {\bibinfo {title} {Quantum Computation and Quantum Information}}}\ (\bibinfo
  {publisher} {Cambridge University Press},\ \bibinfo {address} {Cambridge},\
  \bibinfo {year} {2000})\BibitemShut {NoStop}%
\bibitem [{\citenamefont {Old}\ \emph {et~al.}(2025)\citenamefont {Old},
  \citenamefont {Tasler}, \citenamefont {Hartmann},\ and\ \citenamefont
  {Müller}}]{Old2025fault_tolerant_three_qubit}%
  \BibitemOpen
  \bibfield  {author} {\bibinfo {author} {\bibfnamefont {J.}~\bibnamefont
  {Old}}, \bibinfo {author} {\bibfnamefont {S.}~\bibnamefont {Tasler}},
  \bibinfo {author} {\bibfnamefont {M.~J.}\ \bibnamefont {Hartmann}},\ and\
  \bibinfo {author} {\bibfnamefont {M.}~\bibnamefont {Müller}},\ }\href@noop
  {} {\bibfield  {journal} {\bibinfo  {journal} {arXiv preprint
  arXiv:2506.09029}\ } (\bibinfo {year} {2025})},\ \Eprint
  {https://arxiv.org/abs/2506.09029} {arXiv:2506.09029 [quant‑ph]}
  \BibitemShut {NoStop}%
\bibitem [{\citenamefont {Li}\ and\ \citenamefont
  {Zhang}(2018)}]{li2018onestep}%
  \BibitemOpen
  \bibfield  {author} {\bibinfo {author} {\bibfnamefont {P.-X.}\ \bibnamefont
  {Li}}\ and\ \bibinfo {author} {\bibfnamefont {X.-M.}\ \bibnamefont {Zhang}},\
  }\href {https://doi.org/10.1088/1612-202X/aaa8cc} {\bibfield  {journal}
  {\bibinfo  {journal} {Laser Physics Letters}\ }\textbf {\bibinfo {volume}
  {15}},\ \bibinfo {pages} {035501} (\bibinfo {year} {2018})}\BibitemShut
  {NoStop}%
\bibitem [{\citenamefont {Su}\ \emph {et~al.}(2019)\citenamefont {Su},
  \citenamefont {Liang}, \citenamefont {Gao},\ and\ \citenamefont
  {Zhang}}]{su2019onestep}%
  \BibitemOpen
  \bibfield  {author} {\bibinfo {author} {\bibfnamefont {S.-L.}\ \bibnamefont
  {Su}}, \bibinfo {author} {\bibfnamefont {E.}~\bibnamefont {Liang}}, \bibinfo
  {author} {\bibfnamefont {Y.}~\bibnamefont {Gao}},\ and\ \bibinfo {author}
  {\bibfnamefont {S.}~\bibnamefont {Zhang}},\ }\href
  {https://doi.org/10.1007/s11128-019-2173-6} {\bibfield  {journal} {\bibinfo
  {journal} {Quantum Information Processing}\ }\textbf {\bibinfo {volume}
  {18}},\ \bibinfo {pages} {54} (\bibinfo {year} {2019})}\BibitemShut {NoStop}%
\bibitem [{\citenamefont {Khazali}\ and\ \citenamefont
  {M{\o}lmer}(2020)}]{Khazali2020FastMultiqubit}%
  \BibitemOpen
  \bibfield  {author} {\bibinfo {author} {\bibfnamefont {M.}~\bibnamefont
  {Khazali}}\ and\ \bibinfo {author} {\bibfnamefont {K.}~\bibnamefont
  {M{\o}lmer}},\ }\href {https://doi.org/10.1103/PhysRevX.10.021054} {\bibfield
   {journal} {\bibinfo  {journal} {Physical Review X}\ }\textbf {\bibinfo
  {volume} {10}},\ \bibinfo {pages} {021054} (\bibinfo {year}
  {2020})}\BibitemShut {NoStop}%
\bibitem [{\citenamefont {Ashkarin}\ \emph {et~al.}(2022)\citenamefont
  {Ashkarin}, \citenamefont {Beterov}, \citenamefont {Yakshina}, \citenamefont
  {Tretyakov}, \citenamefont {Entin}, \citenamefont {Ryabtsev}, \citenamefont
  {Cheinet}, \citenamefont {Pham}, \citenamefont {Lepoutre},\ and\
  \citenamefont {Pillet}}]{Ashkarin2022ToffoliFineStructure}%
  \BibitemOpen
  \bibfield  {author} {\bibinfo {author} {\bibfnamefont {I.}~\bibnamefont
  {Ashkarin}}, \bibinfo {author} {\bibfnamefont {I.}~\bibnamefont {Beterov}},
  \bibinfo {author} {\bibfnamefont {E.}~\bibnamefont {Yakshina}}, \bibinfo
  {author} {\bibfnamefont {D.}~\bibnamefont {Tretyakov}}, \bibinfo {author}
  {\bibfnamefont {V.}~\bibnamefont {Entin}}, \bibinfo {author} {\bibfnamefont
  {I.}~\bibnamefont {Ryabtsev}}, \bibinfo {author} {\bibfnamefont
  {P.}~\bibnamefont {Cheinet}}, \bibinfo {author} {\bibfnamefont {K.-L.}\
  \bibnamefont {Pham}}, \bibinfo {author} {\bibfnamefont {S.}~\bibnamefont
  {Lepoutre}},\ and\ \bibinfo {author} {\bibfnamefont {P.}~\bibnamefont
  {Pillet}},\ }\href {https://doi.org/10.1103/PhysRevA.106.032601} {\bibfield
  {journal} {\bibinfo  {journal} {Phys. Rev. A}\ }\textbf {\bibinfo {volume}
  {106}},\ \bibinfo {pages} {032601} (\bibinfo {year} {2022})}\BibitemShut
  {NoStop}%
\bibitem [{\citenamefont {Yu}\ \emph {et~al.}(2022)\citenamefont {Yu},
  \citenamefont {Wang}, \citenamefont {Liu}, \citenamefont {Su}, \citenamefont
  {Qian},\ and\ \citenamefont {Zhang}}]{PhysRevApplied.18.034072}%
  \BibitemOpen
  \bibfield  {author} {\bibinfo {author} {\bibfnamefont {D.}~\bibnamefont
  {Yu}}, \bibinfo {author} {\bibfnamefont {H.}~\bibnamefont {Wang}}, \bibinfo
  {author} {\bibfnamefont {J.-M.}\ \bibnamefont {Liu}}, \bibinfo {author}
  {\bibfnamefont {S.-L.}\ \bibnamefont {Su}}, \bibinfo {author} {\bibfnamefont
  {J.}~\bibnamefont {Qian}},\ and\ \bibinfo {author} {\bibfnamefont
  {W.}~\bibnamefont {Zhang}},\ }\href
  {https://doi.org/10.1103/PhysRevApplied.18.034072} {\bibfield  {journal}
  {\bibinfo  {journal} {Phys. Rev. Applied}\ }\textbf {\bibinfo {volume}
  {18}},\ \bibinfo {pages} {034072} (\bibinfo {year} {2022})}\BibitemShut
  {NoStop}%
\bibitem [{\citenamefont {McDonnell}\ \emph {et~al.}(2022)\citenamefont
  {McDonnell}, \citenamefont {Keary},\ and\ \citenamefont
  {Pritchard}}]{PhysRevLett.129.200501}%
  \BibitemOpen
  \bibfield  {author} {\bibinfo {author} {\bibfnamefont {K.}~\bibnamefont
  {McDonnell}}, \bibinfo {author} {\bibfnamefont {L.~F.}\ \bibnamefont
  {Keary}},\ and\ \bibinfo {author} {\bibfnamefont {J.~D.}\ \bibnamefont
  {Pritchard}},\ }\href {https://doi.org/10.1103/PhysRevLett.129.200501}
  {\bibfield  {journal} {\bibinfo  {journal} {Phys. Rev. Lett.}\ }\textbf
  {\bibinfo {volume} {129}},\ \bibinfo {pages} {200501} (\bibinfo {year}
  {2022})}\BibitemShut {NoStop}%
\bibitem [{\citenamefont {Johansson}\ \emph {et~al.}(2012)\citenamefont
  {Johansson}, \citenamefont {Nation},\ and\ \citenamefont {Nori}}]{qutip2012}%
  \BibitemOpen
  \bibfield  {author} {\bibinfo {author} {\bibfnamefont {J.~R.}\ \bibnamefont
  {Johansson}}, \bibinfo {author} {\bibfnamefont {P.~D.}\ \bibnamefont
  {Nation}},\ and\ \bibinfo {author} {\bibfnamefont {F.}~\bibnamefont {Nori}},\
  }\href {https://doi.org/10.1016/j.cpc.2012.02.021} {\bibfield  {journal}
  {\bibinfo  {journal} {Computer Physics Communications}\ }\textbf {\bibinfo
  {volume} {183}},\ \bibinfo {pages} {1760} (\bibinfo {year}
  {2012})}\BibitemShut {NoStop}%
\bibitem [{\citenamefont {Nielsen}(2002)}]{nielsen2002simple}%
  \BibitemOpen
  \bibfield  {author} {\bibinfo {author} {\bibfnamefont {M.~A.}\ \bibnamefont
  {Nielsen}},\ }\href {https://doi.org/10.1016/S0375-9601(02)01272-0}
  {\bibfield  {journal} {\bibinfo  {journal} {Physics Letters A}\ }\textbf
  {\bibinfo {volume} {303}},\ \bibinfo {pages} {249} (\bibinfo {year}
  {2002})}\BibitemShut {NoStop}%
\bibitem [{\citenamefont {Brinkmann}(2016)}]{brinkmann2016introduction}%
  \BibitemOpen
  \bibfield  {author} {\bibinfo {author} {\bibfnamefont {A.}~\bibnamefont
  {Brinkmann}},\ }\href {https://doi.org/10.1002/cmr.a.21414} {\bibinfo {title}
  {Introduction to average hamiltonian theory. {I}. basics}} (\bibinfo {year}
  {2016})\BibitemShut {NoStop}%
\bibitem [{\citenamefont {Singer}\ \emph {et~al.}(2005)\citenamefont {Singer},
  \citenamefont {Stanojevic}, \citenamefont {Weidemüller},\ and\ \citenamefont
  {Côté}}]{Singer2005}%
  \BibitemOpen
  \bibfield  {author} {\bibinfo {author} {\bibfnamefont {K.}~\bibnamefont
  {Singer}}, \bibinfo {author} {\bibfnamefont {J.}~\bibnamefont {Stanojevic}},
  \bibinfo {author} {\bibfnamefont {M.}~\bibnamefont {Weidemüller}},\ and\
  \bibinfo {author} {\bibfnamefont {R.}~\bibnamefont {Côté}},\ }\href
  {https://doi.org/10.1088/0953-4075/38/2/021} {\bibfield  {journal} {\bibinfo
  {journal} {Journal of Physics B: Atomic, Molecular and Optical Physics}\
  }\textbf {\bibinfo {volume} {38}},\ \bibinfo {pages} {S295} (\bibinfo {year}
  {2005})}\BibitemShut {NoStop}%
\bibitem [{\citenamefont {L{\"o}w}\ \emph {et~al.}(2012)\citenamefont
  {L{\"o}w}, \citenamefont {Weimer}, \citenamefont {Nipper}, \citenamefont
  {Balewski}, \citenamefont {Butscher}, \citenamefont {B{\"u}chler},\ and\
  \citenamefont {Pfau}}]{Loew2012}%
  \BibitemOpen
  \bibfield  {author} {\bibinfo {author} {\bibfnamefont {R.}~\bibnamefont
  {L{\"o}w}}, \bibinfo {author} {\bibfnamefont {H.}~\bibnamefont {Weimer}},
  \bibinfo {author} {\bibfnamefont {J.}~\bibnamefont {Nipper}}, \bibinfo
  {author} {\bibfnamefont {J.~B.}\ \bibnamefont {Balewski}}, \bibinfo {author}
  {\bibfnamefont {B.}~\bibnamefont {Butscher}}, \bibinfo {author}
  {\bibfnamefont {H.~P.}\ \bibnamefont {B{\"u}chler}},\ and\ \bibinfo {author}
  {\bibfnamefont {T.}~\bibnamefont {Pfau}},\ }\href
  {https://doi.org/10.1088/0953-4075/45/11/113001} {\bibfield  {journal}
  {\bibinfo  {journal} {Journal of Physics B: Atomic, Molecular and Optical
  Physics}\ }\textbf {\bibinfo {volume} {45}},\ \bibinfo {pages} {113001}
  (\bibinfo {year} {2012})}\BibitemShut {NoStop}%
\end{thebibliography}%

\end{document}